\documentclass[final,5p,times,twocolumn]{elsarticle}
\usepackage{amsmath}            
\usepackage{epstopdf}           
\usepackage{flushend}    
\usepackage{graphicx}
\usepackage{algpseudocode}
\usepackage{algorithm}
\usepackage{amssymb}
\usepackage{subfigure}

\journal{Elsevier}

\begin{document}

\begin{frontmatter}

\title{Modelling Competitive marketing strategies in Social Networks}

\author[First]{Rahul Goel}
\ead{rahulgoel1106@gmail.com}

\author[First]{Anurag Singh}
\ead{anuragsg@nitdelhi.ac.in}

\author[Third]{Fakhteh Ghanbarnejad}
\ead{fakhteh@pks.mpg.de}

\address[First]{Department of Computer Science and Engineering, National Institute of Technology Delhi, Delhi- 110040, India }
\address[Third]{Institute of Theoretical Physics, Technical University of Berlin,
Hardenbergstr. 36, Sekr. EW 7-1
D-10623 Berlin}

\begin{abstract}
In a competitive marketing, there are a large number of players which produce the same product. Each firm aims to diffuse its product information widely so that it's product will become popular among potential buyers. The more popular is a product of a firm, the higher is the revenue for the firm. A model is developed in which two players compete to spread information in the large network. Players choose their initial seed nodes simultaneously and the information is diffused according to Independent Cascade model (ICM). The main aim of the player is to choose the seed nodes such that they will spread its information to as many nodes as possible in a social network. The rate of spreading of information also plays a very important role in information diffusion process. Any node in a social network will get influenced by none or one or more than one information. We also analyzed how much fraction of nodes in different compartment changes by changing the rate of spreading of information. Finally, a game theory model is developed to obtain the Nash equilibrium based on best response function of the players. This model is based on Hotelling's model of electoral competition.
\end{abstract}


\end{frontmatter}

\section{Introduction}
\label{intro}
A competitive market is one in which a large numbers of producers compete with each other to satisfy the wants and needs of a large number of consumers. In a competitive market no single producer, or group of producers, and no single consumer, or group of consumers, can dictate how the market operates. They can not individually determine the price of goods and services, and how much will be exchanged.

Assume there are two players, $P_1$ and $P_2$, and both are producing the same kind of product. They want to promote their product, which is achieved by spreading information 1 by $P_1$ and information 2 by $P_2$. Both of these informations are competitive. Thus, both firms want to reach large number of consumers. Here, information diffusion comes into play for spreading information.

Information diffusion spreads a piece of information (knowledge) which reaches individuals through interactions \cite{9}. These information can change voting behavior, spread promotion about a product before its launch or increase reputation of a player \cite{1}. Nowadays, many firms keep detailed social information about their customers. These social informations later used by firms for their benefit. In real world, many players spread information simultaneously. By using social networks we can understand this diffusion of information in real world scenarios \cite{diff1, diff2, diff3, 101}. Facebook, Twitter, Google+ are some well-known social platforms.  Social network contains people and they interact with each other. This interaction creates relationship among member of the network. For example, in Facebook networks relationship is friendship, in Google+ network relationship is not just friendship it is a circle. Generally, people get information from their friends or colleagues.

Information diffusion is a vast research domain and can be applied to many fields, such as physics, biology, etc. The diffusion of innovation over network is one of the basic reason for studying networks and spread of diseases among the population. We focus on the particular case of spreading information in social networks, that includes (i) how does information spread in a strategic game environment, (ii) path followed by information into the network, (iii) which node(s) of the network plays an important roles in the spreading process \cite{8}.

Some of the earliest well-structured findings focused in the field of medical and agriculture for the adoption of an idea are described in \cite{103,104}. In contrast of marketing strategies, for the success of product using ``word-of-mouth" and ``viral-marketing", some researches are done in order to see the diffusion process \cite{105,106,107,109,110,111} and the adoption of various strategies in game theory models are given  \cite{112,115,116}. By taking marketing applications, Domingos and Richardson proposed a fundamental algorithmic problem \cite{107,111}.

Nowadays, there are multiple competitive informations and they all target to reach the same people, e.g., companies like Apple, Samsung, etc. produces mobile phones. They are competitor of each other. Their target people are the potential buyer of mobile phones. In order to attract customers they need to promote their product. The company whose product marketing is better than the other, will get more customers and finally more profit. Marketing/promotion requires advertisement, distribution of free samples, word-of-mouth promotion etc. But reaching to every customer personally is not at all possible for any firm. So, the solution is to reach those people who can further promote their product called ``seeds". From seed, information diffusion process starts and reaches to the other potential customer. Hence, there are two important things (1) Choosing the seed node. (2) Information diffusion process.

There are number of papers on competitive information spreading \cite{123,124,125,126}. These competitive information spreadings can easily be understood by game theoretic models. Game theory has become one of the key tools to understand the strategic interactions between individual behavior. Game theory and its main concepts like Nash equilibrium can be used for both cooperative and competitive behavior \cite{127,128}. Until now, it is applied in almost every field. Some of these fields include computer science, biology, ecology, sociology, public health, traffic management, economics, and mathematics \cite{129,132,133,134}.

We studied Facebook dataset, which helps to understand the information diffusion process in real world scenario. Information starts from some initial nodes and reaches to the entire network. This can be easily seen as an example of product marketing. Other dataset used is Wiki-vote in which users are voting to promote user to adminship. 

We applied three methods (1) Degree centrality [consider node local property], (2) Eigenvector centrality [consider node global property] and (3) Rank Degree method [use sampling of the network], for seed selection. These methods allow us to choose very different types of seed nodes which are important in the network in one way or the other. For information spreading, independent cascade model (ICM) with threshold is implemented. It is one of the popular methods to analyze the effect on node due to its neighboring nodes.

In 1978, Mark Granovetter \cite{cml1}, developed threshold models for collective behavior, based on behavioral threshold. The analysis is done on binary decision of actor/nodes in a network. Beginning with a frequency distribution of thresholds, the models allow calculation of the ultimate or ``equilibrium" number making each decision. The stability of equilibrium results against various possible changes in threshold distribution is also considered. 

In 2002, Duncan J. Watts \cite{cml2}, investigated the global cascades in random networks. They showed that global cascades in social and economic systems, as well as cascading failures in networks occur rarely. But gloabal cascades and cascading failures are large when they occur. They studied binary-decision model in different conditions. When the network of interpersonal influence is sufficiently sparse, the propagation of cascades is limited by the global connectivity of the network; and when it is sufficiently dense, cascade propagation is limited by the stability of the individual nodes. In first case, distribution of cascade is power-law and in the other case, it is bimodal. Increased heterogeneity of individual thresholds appears to increase the likelihood of global cascades; but increased heterogeneity of vertex degree appears to reduce it.

D. Centola et al. \cite{cml3}, studied the cascade dynamics of multiplex propagation. They showed that random links between otherwise distant nodes can greatly facilitate the propagation of disease or information, provided contagion can be transmitted by a single active node. However, when the propagation requires simultaneous exposure to multiple sources of activation, called multiplex propagation, the effect of random links makes the propagation more difficult to achieve.

M. Karsai et al. \cite{cml4}, studied the effect of different topological and temporal correlations on spreading in complex communication networks. They showed that (i) the community structure and its correlation with link weights and (ii) the inhomogeneous and bursty activity patterns on the links, plays an important role in spreading speed.

F. Karimi and P. Holme \cite{cml5}, studied threshold models of cascade in a temporal model by extending Watts' cascade model \cite{cml2}. They assumed that people are influenced by their past contacts. Two versions of the model are investigated, where they respond to the absolute number of such contacts, They observe that temporal network structure heavily affects these models. In fractional threshold model, the cascade size decreases with time window size, but the size of cascade is large for randomization. In absolute threshold mode, the situation is the opposite.

The seed selection methods and ICM helps us to see insight of marketing. The ways by which, a small number of initial nodes can affect the entire network. The dynamics of this diffusion can be understood by ICM.

In the proposed model, two players are considered with two competing information respectively. Each player has a fixed initial budget, which he can spend to select seed node(s). Players choose their seed nodes simultaneously and then information diffuses according to ICM \cite{7}. ICM represent a network using directed graph \cite{9} and a node can inform another node in the network in its acquaintance. Therefore, nodes may or may not be informed in this model. In the network, nodes which are informed by at least one information is considered as informed nodes for the same information. In the underlying network each node has a threshold for informing itself by a particular information. If affect of that information on the node is greater or equal to threshold value then it will be considered as informed. A node once informed can further inform its neighboring nodes. This is a progressive process, where nodes change from non-informed to informed, but not vice versa.

If a node is informed by more than one informations then, the information which will affect more is considered as supporter of that information. If node is effected equally by both informations, then assign that node as a supporter for only one of the information with equal probability.

The proposed model initially converts network into tree using seed nodes as the root node. Influence of sibling nodes is also considered for each node at that level of the tree. This gives more detailed influence of information for a node. We have also considered node cost in order to choose seed node. The proposed model is able to find influential spreader in social networks for both firms. Later, it spreads both firms information so that maximum number of nodes in the network are effected and at the same time, equilibrium is achieved in terms of supporter for both firms.

In section 2, we describe the proposed model, choosing spreader node and information spreading using Independent Cascade model. In Section 3, Mean field approximation is explained. In section 4, game theory model is explained. In section 5, Simulation and Results are shown. In section 6, Conclusions are given.

\section{Proposed Model}
\label{sec:model}
Let $G(V,E)$ be an unweighted and undirected network with a set of $n$ nodes $V:=\left\{1,2,...,n\right\}$ and a set of links $E$. We denote neighbors of $i \in V$ as $N_i(G):=\left\{j|(j,i)\in E\right\}$ and degree of $i$ as $d_i:=|N_i(G)|$. A threshold (how much a node is informed by any information) for node $i$, denoted as $\theta_i$ is a probability between $[0, 1]$. Node with influence of information(s) greater than or equal to the threshold, is considered as affected by that information(s). The influence of information is defined as the effect on the behavior of node due to the information. Its value lies between 0 to 1, where, 0 means no influence and 1 means complete influence of information.

There are two players, $P_1$ and $P_2$ and two informations 1 and 2, respectively. In a network, a node can be Informed/spreader (means node is informed by atleast one information), supporter (means node is supporting one of the information) and non-spreader/uninformed (means no information reached to the node or information(s) reached to the node but its influence is less than threshold of the node). Nodes on a social network either support information 1 or information 2 or remain non-spreader.

\begin{figure}[h!]
\centering
\includegraphics[width=0.7\linewidth, height=1.8 in]{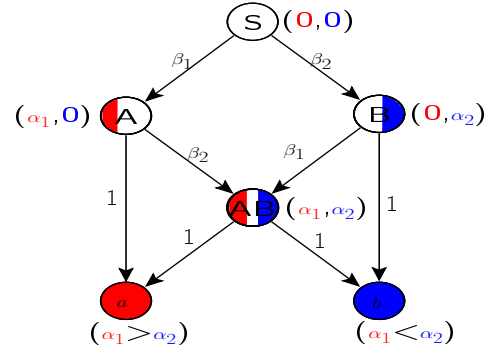}
\caption{Flow chart for state of nodes, in a competitive environment.}
\label{flow}
\end{figure}

In the proposed model, six compartments are possible for each node, \textbf{S}, \textbf{A}, \textbf{B}, \textbf{AB}, $a$ and $b$ (Figure ~\ref{flow}). All these variables (\textbf{S}, \textbf{A}, \textbf{B}, \textbf{AB}, $a$ and $b$) are stochastic variables and sum of all these compartment is 1 at any time. Here, \textbf{S} represents compartment for uninformed nodes, \textbf{A} and \textbf{B} represent compartments for informed nodes by information 1 and information 2, respectively and \textbf{AB} represents compartment for nodes informed by both information 1 and 2, while $a$ and $b$ represents compartments for supporter of information 1 and 2, respectively. For example, a node in compartment \textbf{A} is informed by information 1, a node in compartment \textbf{AB} is informed by both, information 1 and information 2, a node in compartment $a$ is supporter of information 1. Let $\alpha_1$ and $\alpha_2$ be the influence of information 1 and 2, respectively ($0\leq\alpha_1\leq 1;0\leq\alpha_2\leq 1$). The values of $\alpha_1$ and $\alpha_2$ are most probably different for different nodes in the network. For better understanding, different colors are used, red for information 1 and blue for information 2. For example, a node in compartment \textbf{AB} is informed by information 1 with value $\alpha_1$ as shown in red. Similarly, influence of information 2 with value $\alpha_2$ as shown in blue. A node in compartment $a$ is supporter of information 1. Hence, it is shown completely in red color. Same is observed for node in compartment $b$ which supports information 2 and shown in blue color only. (\textbf{Note}: For a node if ($\alpha_1==\alpha_2$) then, it supports both information with probability 0.5). $\beta_1$ and $\beta_2$ represent rate of spreading of information 1 and information 2.

Let, $\alpha^{par}_{i}$ be the influence of information $i$ (where, $i$ = 1, 2) on parent node, $\alpha^{ch}_{i}$ be the influence of information $i$ on child node, $d_{ch}$ be the degree of child node, $\kappa$ be the set of children of $par$, $A$ be the adjacency matrix and $\alpha^{sib}_{i}$ be the influence of information $i$ on node $sib$ of parent node only, where, $sib\in \kappa$. So, influence of information $i$ on node $ch$ is given as

\begin{align}
\alpha^{ch}_i&=\sum_{par\in\kappa}\left(\frac{\alpha^{par}_{i}}{d_{ch}}.A(ch,par)\right)+\sum_{sib\in\kappa}\left(\frac{\alpha^{sib}_{i}}{d_{ch}}.A(ch,sib)\right)
\end{align}
\begin{align*}
ch,par,sib&\in V
\end{align*}

Initially, all nodes except seed nodes are in uninformed compartment. Players choose seed node simultaneously for their information. A player can use a node as a seed,  if it has budget to get that node. The proposed method is described in the algorithm \ref{proposed model} with time complexity of $O(VE)$.

\scriptsize
\begin{center}
 \begin{tabular}{||c | c||}
 \hline
 Notation & Meaning\\ [0.5ex] 
 \hline\hline
 \textbf{$V$} & Number of vertices\\
 \hline
\textbf{$E$} & Number of edges\\
\hline
\textbf{$d_i$}& Degree of node $`i'$\\
\hline
\textbf{$\theta_i$} & Threshold for node $`i'$\\
\hline
\textbf{$P_1,\ P_2$} & Players\\
\hline
$\mathbf{S}$ & Uninformed nodes\\
\hline
$\mathbf{A}$ & Informed nodes by information 1\\
\hline
$\mathbf{B}$ & Informed nodes by information 2\\
\hline
$\mathbf{AB}$ & Informed nodes by information 1 and 2 both\\
\hline
\textbf{$a$} & Supporter nodes of information 1\\\hline
\textbf{$b$} & Supporter nodes of information 2\\\hline
\textbf{$\alpha_1,\ \alpha_2$} & Influence of information\\\hline
\textbf{$\beta_1,\ \beta_2$} & Rate of spreading of information\\\hline
\textbf{$\alpha_{i}^{par}$} & Influence of information $`i'$ on parent node\\\hline
\textbf{$\alpha_{i}^{ch}$} & Influence of information $`i'$ on child node\\\hline
\textbf{$\kappa$} & Set of children of parent $`par'$\\\hline
\textbf{$Adj$} & Network adjacency matrix\\\hline
\textbf{$\alpha_{i}^{sib}$} & Influence of information $`i'$ on sibling node of parent node only\\\hline
\textbf{$c_i$} & Cost of node $`i'$\\\hline
\textbf{$d_{ct}$} & Degree of central tendency node\\\hline
\textbf{$B_i$} & Budget of player\\\hline
\textbf{$<k>$} & Average degree\\\hline
\textbf{$C.C.$} & Clustering Coefficient $`i'$\\\hline
\textbf{$\mu_{influenced}$} & Fraction of influenced nodes\\\hline
\textbf{$\mu_{supporter}$} & Fraction of supporter nodes\\\hline
\textbf{$L$} & Number of levels\\\hline
\textbf{$F_1,\ F_2$} & Firms\\\hline
\textbf{$x_1,\ x_2$} & Position of firms based on informed nodes\\\hline
\textbf{$B_i(x_j)$} & Best response function of firm $`i'$ when firm $`j'$ position is $x_j$\\\hline
\textbf{$y_i,\ y_2$} & Position of firms based on supporter nodes\\\hline
\textbf{$B_i(y_j)$} & Best response function of firm $`i'$ when firm $`j'$ position is $y_j$\\ [1ex] 
 \hline
\end{tabular}
\end{center}

\normalsize

\begin{algorithm}
\caption{\textbf{: Proposed Method}}
\label{proposed model}
\begin{algorithmic}[1]
\State Calculate the cost for each node in a network
\State Initialize the budget for each player
\State Choose seed node(s) for each player simultaneously
\State Spread information with Independent Cascade model and identify supporter(s) for each information
\State Repeat step 2 to 5, multiple times
\end{algorithmic}
\end{algorithm}

\subsection{Calculating cost for each node}
\label{sec:Calculating cost for each node}
Node cost is the price of node to select as a seed node. This cost is paid by the firm from the allocated budget. We are using the central tendency concept to decide the cost for nodes. A central tendency (or measure of central tendency) is a central or typical value for a probability distribution \cite{central}. It may also be called a center or location of the distribution. Colloquially, measures of central tendency are often called averages.
The most common measures of central tendency are the arithmetic mean, the median and the mode. 

There are two kinds of outliers in any data: Bad outlier and Good outlier. \textbf{Bad Outlier}: is at an observation that lies at abnormal distance from other values in a random sample from a population. In this case, we use median. \textbf{Good Outlier}: is an observation that lies at a normal distance from other values in a random sample from a population. In this case, we use mean.

We are considering the degree of a node in order to calculate its cost. The degree of nodes contains bad outliers, so we are using median as central tendency. Assign unit cost to the nodes with central tendency. Cost of remaining nodes can be decided using linear method.

Node cost estimation is given in algorithm \ref{node cost}. Time complexity for finding central tendency degree using median takes $O(n)$, where $n$ is total number of nodes in the network. Assignment of cost using linear method also takes time $O(n)$. Hence, time complexity for algorithm node cost estimation is $O(n)$.

\begin{algorithm}
\caption{\textbf{: Node Cost estimation}}
\label{node cost}
\begin{algorithmic}[1]
\State Find the central tendency degree using Median.
\State Assign all nodes with degree equal to central tendency degree as unit cost.
\State Calculate cost for other nodes using linear method.
\end{algorithmic}
\end{algorithm}

Let $G(V,E)$ be an undirected network with $V$ nodes and $E$ edges. Degree of any node $i$ is $d_i$ and degree of central tendency node is $d_{ct}$. Cost of node $i$ is $c_i$. Hence,

\textbf{Linear Method for calculating cost}:
\begin{equation}
\forall i \hspace{2 mm} c_i= \frac{d_i}{d_{ct}}, \hspace{5 mm} i=1,2,...,V
\end{equation}
\subsection{Initializing the Budget for Player}
\label{sec:Initializing the Budget}
Each player, $P_i$, is initialized with a unit budget $B_i(B_i=1)$ to be spent for choosing seed. The idea is that, initially player must be able to choose at most the node with degree as central tendency as a seed node. And based on the goodness of the seed node, player will win or lose from other firm. We can change the initial budget for players according to our need. Let consider there are $f$ number of firms then time complexity for initializing budget is $O(f)$, where, $f<<<n$.
\subsection{Choosing Spreader Node}
There are many methods to decide the importance of a node. Methods we used to choose seed are:
\begin{enumerate}
\item Degree Centrality (DC).
\item Eigenvector Centrality (EC).
\item Rank Degree (RD).
\end{enumerate}

This list of methods is not exhaustive. Hence, in the initial phase we aimed for these methods.
\subsubsection{Degree Centrality}
Degree centrality is the simplest index to identify nodes influences. In case of more connections, a node gets greater influence. To compare the influences of nodes in different networks, the normalized degree centrality is defined as 
\begin{equation}
DC(i)=\frac{d_i}{n-1}
\end{equation}
where, $n=|V|$ is the number of nodes in $G$ and $n$-1 is the largest possible degree \cite{2}.

For dense adjacency matrix representation of the graph, calculating degree centrality for all the nodes in a graph takes ${\displaystyle \Theta (V^{2})}$. For sparse matrix representation of the graph, calculating degree centrality for all the nodes in a graph takes ${\displaystyle \Theta (E)}$.

\subsubsection{Eigenvector Centrality}
Eigenvector centrality supposes that the influence of a node is not only determined by its neighbors, but also determined by the influence of each neighbor \cite{2,4}.The centrality of a node is proportional to the summation of the centralities of the nodes to which it is connected. The importance of a node $i$, denoted by $\chi_i$ is
\begin{equation}
\chi_i=c\sum_{j=1}^{n} a_{ij}\chi_j,
\end{equation}

\begin{algorithm}
\caption{Eigenvector Centrality}
\label{eigenvector}
\begin{algorithmic}[1]
\State Start by	assigning centrality score of 1 to all nodes($\chi_i$=1, $\forall$ $i$ in the network)	
\State Recompute scores of each node as weighted sum of centralities of all nodes in a node's neighborhood:$$\chi_i=c\sum_{j\in N} a_{ij}\chi_j$$
\State Normalize $\chi$ by dividing each value by the largest value
\State Repeat steps 2 and 3 until values of $\chi$ converge.
\end{algorithmic}
\end{algorithm}
which can be written in the matrix form as
\begin{equation}
{\chi}=cA\overrightarrow{\chi}
\end{equation}
where, $c$ is a proportionality constant. Generally, $c=1/\lambda$ in which $\lambda$ is the largest eigenvalue of $A$. Time complexity for eigenvector centrality is $O(V^3)$.
\subsubsection{Rank Degree}
This method is based on graph sampling, the problem of selecting a small subgraph which has the topological properties as the original graph. A sampling method effectively identifies the influential spreader if and only if (a) the fraction of top-k common nodes in the samples and in the graph is on an average sufficiently large and (b) the ranking of these nodes in the samples are close to the original ranking in the graph \cite{6}. Time complexity of rank degree algorithm is $O(n^2)$ for sparse matrix.

\begin{algorithm}
\caption{Rank Degree Algorithm}
\label{rankdegree}
\begin{algorithmic}[1]
\State \textbf{Set parameters:} (i)s:number of initial seeds, (ii) $\rho$, (iii) target sample size $x$
\State \textbf{Input:} undirected graph $G(V,E)$
\State \textbf{Output:} sample of size $x$
\State \textbf{Initialization:} $\left\{Seeds\right\} \gets s$ nodes selected uniformly at random
\State $Sample \gets \phi$
\While{sample size $<$ target size $x$}
\State $\left\{New \hspace{1 mm} Seeds\right\} \gets \phi$
\For{$\forall w\in \left\{Seeds\right\}$ }
\State Rank $w's$ friends based on their degree values
\State \textbf{Selection Rule:}
\State (i)$RD(max)\in$select the max degree (top-1) friends of $w$
\State (ii)$RD(\rho)\in$select the top-k friends of w, where $k=\rho.(\#friends(w)),$ 0$<\rho\leq$1
\State Update the current sample with the selected edges $(w,friend(w)$ on the top-k) along with the symmetric ones
\State Add to $\left\{New\ Seeds\right\}$ the top-k friends of $w$
\EndFor
\State Update graph G: delete from the graph all the currently selected edges
\State $\left\{Seeds\right\}\gets\left\{New\ Seeds\right\}$
\If{$\left\{New\ Seeds\right\}=\phi$} repeat Step-4 (random jump)
\EndIf
\EndWhile
\end{algorithmic}
\end{algorithm}

\subsection{Information Spreading}
\label{InformationSpreading}
Nodes chosen as initial spreaders are assigned values of $\alpha_1$ and $\alpha_2$ as 1. After choosing the seed node(s), we use independent cascade to see how many nodes are informed by each information. The spreading process involves the three basic elements: Sender, Receiver and Medium.
\subsubsection{Cascade Model}
Information Cascade is defined as \textit{``A  behavior of information adoption by people in a social network resulting from the fact that people ignore their own information signals and make decisions from inferences based on earlier people's actions"} \cite{8}.

In Cascade spreading, network is converted into directed tree(s) using seed node(s) as a root for each information. Using these trees, we find the influence of each information on nodes level by level. The influence of siblings, nodes with the same parents, at each level is also considered. If there is an edge exist between siblings in the original network.

The diffusion process is characterized by two aspects: its structure, i.e., the diffusion graph that who influenced whom, and its temporal dynamics, i.e., a number of nodes that adopts the piece of information over time. The simplest way to describe the spreading process is to consider that a node can be either spreader (i.e., has received the information and tries to propagate it) or non-spreader.

A sample network as shown in Figure \ref{fig:icm1}(a) is taken in order to explain the information spreading process using ICM. Here, node 2 is the seed node for the information 1 and node 5 is the seed node for the information 2. Sample network is treated as tree like structure by both seeds to spread their respective information simultaneously. Seeds node are the origin point of information hence, considered as the root node for their information. Figure \ref{fig:icm1}(b) shows the tree-like structure of the sample network by taking node 2 as a seed node for the information 1. Similarly, for the same sample network, tree-like structure by taking node 5 as the seed node for the information 2 is shown in Figure \ref{fig:icm1}(d). There is only the network in the reality, not any tree but we are considering tree to make it convenient to explain the spreading process.

Once the selection of seed nodes is done, both seeds spread their information simultaneously. Information will spread level by level according to the tree like structure with respect to the seed node for the information. The link/connection between the siblings is shown by using colored two-way arrows. This indicates that siblings will pass information to each other. Hence, the influence of information increases on both of the siblings.

Figure \ref{fig:icm1}(b) shows tree like structure for seed (node 2) of the information 1 and this information will not be able to flow beyond seed (node 5) of the information 2, so we simply removed seed of the information 2 and all the other nodes not reachable from node 2. The resultant tree will look like as shown in Figure \ref{fig:icm1}(c). Now, calculate the influence of information 1 spread by seed (node 2) on each node of the network. Similarly, the influence of information 2 spread by seed (node 5) on each node of the same network is shown in Figure \ref{fig:icm1}(d) and (e). Finally, from Figure \ref{fig:icm1}(c) and (e) we can easily check influence of the information 1 and the information 2 on the sample network as shown in the Figure \ref{icm4}(a).

\begin{figure}[ht!]
\centering
\includegraphics[width=0.8\linewidth, height=1.7 in]{Figures/icm1.eps}\\
\mbox{\textbf{(a)} }
$\begin{array}{cc}
\includegraphics[width=0.45\linewidth, height=1.3 in]{Figures/icm2.eps} &
\includegraphics[width=0.45\linewidth, height=1.3 in]{Figures/icm3_1.eps}\\
\mbox{\textbf{(b)} } & \mbox {\textbf{(c)}}
\end{array}$
$\begin{array}{cc}
\includegraphics[width=0.45\linewidth, height=1.3 in]{Figures/icm3.eps} &
\includegraphics[width=0.45\linewidth, height=1.3 in]{Figures/icm2_1.eps}\\
\mbox{\textbf{(d)} } & \mbox {\textbf{(e)}}
\end{array}$
\caption{Sample Network with node 2 and node 5 as the seed for information 1 and 2, respectively.}
\label{fig:icm1}
\vspace{-1em}
\end{figure}

In a competitive environment with two firms $F_1$ and $F_2$, there are three cases possible, (i) Firm $F_1$ wins, (ii) Firm $F_2$ wins, (iii) Tie. As we can see in Figure \ref{icm4}(a), node 2 is the seed for information 1 and node 5 are the seed for information 2. Influence of information 1 on nodes 1, 2, 3, 4 and 10 is higher than information 2. Hence, they support information 1. Similarly, the influence of information 2 on nodes 5, 6, 7, 8 and 9 are higher than information 1. Hence, they support information 2. As there are equal numbers of supporters for both informations, this is a tie case. For detailed analysis refer Appendix B. This process can further be extended to any number of players or firms.

Figure \ref{icm4}(b) shows the case where firm $F_1$ wins. Node 3 is the seed and nodes 1, 2, 3, 4 and 10 are supporters for information 1. Node 5 is the seed and nodes 5 and 6 are supporters for information 2. Here, nodes 7, 8 and 9 are equally influenced by both information. Hence, they support both informations with equal probability.
By comparing values of influence of information at each node, we can decide the polarity of that node. Time complexity for information spreading is $O(VE)$.

\begin{figure}[htb!]
\begin{center}
$\begin{array}{cc}
\includegraphics[width=0.45\linewidth, height=1.3 in]{Figures/icm4.eps} &
\includegraphics[width=0.45\linewidth, height=1.3 in]{Figures/icm5.eps}\\
\mbox{\textbf{(a) Tie} } & \mbox {\textbf{(b) A wins}}
\end{array}$
\end{center}
\caption{Spreading information in the network.} \label{icm4}
\vspace{-1em}
\end{figure}

\section{Mean Field Approximation}

Why rate of spreading of information(s) is important in a competitive environment ? How behavior of population changes with the change in the rate of spreading of information(s) ? This section answers these questions by approximating the spreading dynamics in mean field framework (Figure \ref{flow}).

In the proposed model, six compartments are possible for each node, \textbf{S, A, B, AB}, $a$ and $b$. All these variables are stochastic variables and sum of all these compartment is 1 at any time. $\beta_1$ and $\beta_2$ are rate of spreading of information and independent of each other. A node in compartment \textbf{S} can change its compartment to compartment \textbf{A} or \textbf{B} depending upon the rate of spreading of information, $\beta_1$ and $\beta_2$. If $\beta_1>\beta_2$, then there are greater chances that node will change its compartment from \textbf{S} to \textbf{A} than \textbf{A} to \textbf{B}. Similarly, if $\beta_1<\beta_2$, then there are less chances that node will change its compartment from \textbf{S} to \textbf{A} than \textbf{A} to \textbf{B}. If $\beta_1=\beta_2$, nodes in state \textbf{S} have equal chances to change its compartment form \textbf{S} to \textbf{A} and \textbf{S} to \textbf{B}. 

If a node is informed by any of the information, then, it may get informed by other information. Chances of the node to move from compartment \textbf{A} or \textbf{B} again depends upon the rate of spreading of other information. If rate of spreading of other information is sufficiently high to influence the node, then it changes its compartment from \textbf{A} or \textbf{B} to \textbf{AB}. If node is informed by both informations, then, the node changes its compartment from \textbf{AB} to $a$ or $b$ depending upon influence of information on the node. Otherwise, the node is only informed by one of the information, so, it changes its compartment from \textbf{A} to $a$ or \textbf{B} to $b$.

On the basis of model discussed, ODE's are formulated and are given as follows:

\begin{equation}
\begin{split}
[\dot{S}] &= -\beta_1[S]([A] + [AB]) - \beta_2[S]([B]+[AB]) \\
[\dot{A}] &= \beta_1[S]([A]+[AB])-\beta_2[A]([B]+[AB])-[A] \\
[\dot{B}] &= \beta_2[S]([B]+[AB])-\beta_1[B]([A]+[AB])-[B] \\
[\dot{AB}] &= \beta_2[A]([B]+[AB])+\beta_1[B]([A]+[AB])-2[AB] \\
[\dot{a}] &= [A] + [AB] \\
[\dot{b}] &= [B] + [AB]
\end{split}
\end{equation} 

\begin{equation}
[S]+[A]+[B]+[AB]+[a]+[b]=1
\end{equation}

\textbf{Note:} For detailed analysis refer Appendix A.

\section{Game Theory Model}
During product launch, firms aim to reach a large fraction of population using campaign. The free sample distribution is one of the popular campaign strategies. How free samples of the newly launched product to be distributed among the population so that the selected people will influence the large fraction of potential customers? During the election, each candidate has its own policies to win. Basically, policies are made for targeting voters. So, these policies must reach to all voters in less time and with high influence. Whom to target during the election in order to reach maximum people with candidate policies? Whom to select your brand ambassador, so that influence of your product is high on a large number of the population ? The model is a foundation for answering many of the such questions, i.e., Electoral competition or Hotelling's model of electoral competition \cite{GT1,GT2}. We are using Hotelling's model in our proposed work for solving the game theoretic problem.

This problem is a strategic game in which the players are the firms and seed selection, referred to as a ``position". This position reflects the importance of seed in the network, i.e., how much fraction of nodes are informed, spread by the seed node. Once firms choose their seed node(s), their position is fixed and can be determined by using Information spreading process (Section \ref{InformationSpreading}).

This position is dependent on the cost of the node(s) and budget of firms. Node cost estimation is given in section \ref{sec:Calculating cost for each node} and budget for firms is described in section \ref{sec:Initializing the Budget}. If a firm can afford to pay a high cost for the seed node, then, there is a higher probability that firm position is better than other firm. So, the fraction of informed and supporter nodes for that firm will be higher. It assures that budget of the firm plays an important role for the information spreading process in the competitive environment.

\subsection{Position of firms on the basis of proportion of informed nodes}
\begin{figure}[ht!]
\centering
\includegraphics[width=\linewidth, height=1.6 in]{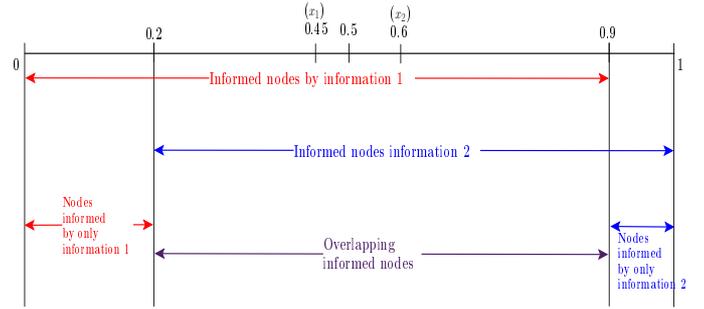}
\caption{Position of firms on the basis of proportion of informed nodes}
\label{fig:posinf}
\end{figure}

In this section, we discuss the position of the firms on the basis of informed nodes. Suppose 0.9 proportion of nodes of the population is informed by information 1 and 0.8 proportion of nodes of the population is informed by information 2. Then, position of firms on the basis of the informed node is given in Figure \ref{fig:posinf}. Here, $x_1$ and $x_2$ are positions of firm 1 and firm 2 respectively. A node can be informed by information 1 or information 2 or, both.

Depending upon the position of firms on the basis of informed nodes, for a firm $i$, the whole population of the network can be broadly divided into two parts, (I) Uninformed nodes by information $i$ and (II) Informed nodes by information $i$. Informed nodes by information $i$ can further be divided into (a) nodes informed by only information $i$ and (b) nodes informed by both information $i$ and information $j$ (This is also called overlapping proportion of the population of nodes).

Population size is considered large. For convenience, we normalize the population of nodes from 0 to 1. The distribution of nodes is uniform. The mean position: the position 0.5 with the property that exactly half population of the network lies to the left of this position and other half population lies to the right of this position.

Once firm(s) found proportion of population of informed by information of that firm(s), then we can set interval for those firm(s) also. In Figure \ref{fig:posinf}, firm 1's informed nodes are 0.9, its interval is from 0 to 0.9 (and position of firm 1 is 0.45 $\left[\frac{0.9-0.0}{2}\right]$) $(0.0\leq x_1\leq 0.5)$. Firm 2's informed nodes are 0.8, its interval is from 0.2 (1.0-0.8) to 1.0 (and position of firm 2 is 0.6 $\left[\frac{1.0+(1.0-0.8)}{2}\right]$) $(0.5\leq x_2\leq 1.0)$.

The proposed problem can be formulated as game theoretic model as given below
\begin{itemize}
\item $Players$\hspace{1mm} The Firms.
\item $Actions$\hspace{1mm} Each firm's set of actions is the set of positions.
\item $Preferences$\hspace{1mm} Each firm's preferences are represented by payoff function.
\end{itemize}

Suppose there are two firms. We can find a Nash equilibrium of the game by the player's best response functions. Fix the position of $x_2$ of firm 2 and consider the best position for firm 1. First, suppose $x_2>$0.5. If firm 1 takes a position from (1-$x_2$) to 0.5 [(1-$x_2)<x_1\leq$0.5] then, firm 1 wins and firm 2 loses. If firm 1 takes a position less than (1-$x_2$) [$x_1<(1-x_2$)] then, firm 1 loses and firm wins. If firm 1 takes position exactly at (1-$x_2$) [$x_1=(1-x_2)$] then, tie for both firms.

Firm 1's best response function is defined as 
\begin{equation}
  B_1(x_2)=\begin{cases}
    \left\{x_1:(1-x_2)<x_1\leq 0.5\right\}, & if\ x_2>0.5.\\
    \{0.5\}, & if\ x_2=0.5.
  \end{cases}
\end{equation}

Similarly, Firm 2's best response function is defined as 
\begin{equation}
  B_2(x_1)=\begin{cases}
    \left\{x_2:0.5\leq x_2<(1-x_1)\right\}, & if\ x_1<0.5.\\
    \{0.5\}, & if\ x_1=0.5.
  \end{cases}
\end{equation}

\begin{figure}[ht!]
\centering
\includegraphics[width=0.7\linewidth, height=1.5 in]{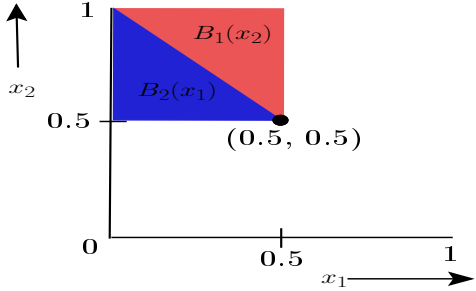}
\caption{Nash Equilibrium for Position of firms on the basis of informed nodes}
\label{fig:eq1}
\end{figure}

The firms' best response functions are shown in Figure \ref{fig:brf1}.
\begin{figure}[ht!]
\begin{center}
$\begin{array}{cc}
\includegraphics[width=0.45\linewidth, height=1.3 in]{Figures/Diagram2.eps} &
\includegraphics[width=0.45\linewidth, height=1.3 in]{Figures/Diagram3.eps} \\
\end{array}$
\caption{Best response function for firm 1 and firm 2.}
\label{fig:brf1}
\end{center}
\end{figure}
If you superimpose the two best response functions, you see that the game has
a unique Nash equilibrium, in which both firms choose the position 0.5, the mean position. We can make an argument that (0.5, 0.5) is the unique Nash equilibrium of the game as shown in Figure \ref{fig:eq1}. We can also argue that First, (0.5, 0.5) is an equilibrium: it results in a tie, and if either firm chooses a position different from 0.5 then, it loses. Second, no other pair of positions is a Nash equilibrium, by the following argument.

(1) If one firm loses then, it can do better by moving to 0.5, where it either wins outright (if other firm's position is different from 0.5) or ties
for first place (if other firm's position is 0.5).
(2) If the firms tie (because their positions are symmetric about 0.5) then, either firm can do better by moving to 0.5, where it wins outright.

\subsection{Position of firms on the basis of proportion of supporter nodes}
\begin{figure}[ht!]
\centering
\includegraphics[width=0.8\linewidth, height=1.5 in]{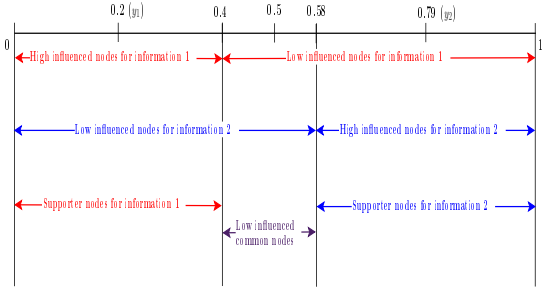}
\caption{Position of firms on the basis of proportion of supporter nodes}
\label{fig:possupp}
\end{figure}

In this section, we discuss the position of the firms on the basis of supporter nodes. Suppose, 0.4 proportion of nodes of the population are the supporter of information 1 and 0.42 proportion of nodes of the population are the supporter of information 2. Then, firms position on the basis of supporter node is given in Figure \ref{fig:possupp}. Here, $y_1$ and $y_2$ are positions of firm 1 and firm 2, respectively.

Depending upon the position of firms on the basis of supporter nodes for a firm $i$, the whole population of the network can be broadly divided into two parts, i.e., first, High influenced nodes by information $i$ and second, Low influenced nodes by information $i$. High influence nodes by information $i$ are those nodes which support information $i$ over information $j$. So, these nodes are more influenced by information $i$ than information $j$. Low influenced nodes for information $i$ contain nodes which are not informed by information $i$ and nodes informed by information $i$ but the influence of information $i$ is less than the threshold of the nodes (threshold, $\theta$ for nodes is described in section \ref{sec:model}).

Low influenced common proportion of nodes are those nodes which support no information. This behavior is observed because no information reached these nodes or information(s) reached to these proportion of nodes but the influence of those information(s) is less than the threshold of the node.

There may be nodes which are equally influenced by both information. In this case, they support both information with equal probability. Hence, considered supporter for both information.

The firm(s) who obtains the most proportion of supporter nodes gets more profit. Each firm cares only about profit; no firm has an ideological attachment to any position. Specifically, each firm prefers to high profit than to tie (equal profit for both firm) for the first place than to low profit, and if it ties for the first place it prefers to do so with as few other firms as possible.

Once fraction of supporter nodes is found for the firm(s) then, we can set interval for those firm(s) also. Example, In Figure \ref{fig:possupp}, firm 1's supporter nodes are 0.4, its interval is from 0 to 0.4 (and position of firm 1 is 0.2 $\left[\frac{0.4-0.0}{2}\right]$) $(0.0\leq y_1\leq 0.5)$. Firm 2's supporter nodes are 0.42, its interval is from 0.58 (1.0-0.42) to 1.0 (and position of firm 2 is 0.79 $\left[\frac{1.0+(1.0-0.42)}{2}\right]$) $(0.5\leq y_2\leq 1.0)$.

Similarly, here we can find a Nash equilibrium of the game by the player's best response functions shown in Figure \ref{fig:brf2}. 

Firm 1's best response function is defined as 
\begin{equation}
  B_1(y_2)=\begin{cases}
    \left\{y_1:(1-y_2)<y_1\leq 0.5\right\}, & if\ y_2>0.5.\\
    \{0.5\}, & if\ y_2=0.5.
  \end{cases}
\end{equation}

Similarly, Firm 2's best response function is defined as 
\begin{equation}
  B_2(y_1)=\begin{cases}
    \left\{y_2:0.5\leq y_2<(1-y_1)\right\}, & if\ y_1<0.5.\\
    \{0.5\}, & if\ y_1=0.5.
  \end{cases}
\end{equation}

\begin{figure}[ht!]
\begin{center}
$\begin{array}{cc}
\includegraphics[width=0.45\linewidth, height=1.5 inn]{Figures/Diagram5.eps} &
\includegraphics[width=0.45\linewidth, height=1.5 in]{Figures/Diagram6.eps} \\
\end{array}$
\caption{Best response function for firm 1 and firm 2.}
\label{fig:brf2}
\end{center}
\end{figure}

\begin{figure}[ht!]
\centering
\includegraphics[width=0.7\linewidth, height=1.5 in]{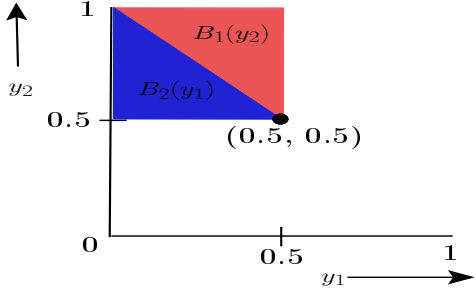}
\caption{Nash Equilibrium for Position of firms on the basis of supporter nodes}
\label{fig:eq2}
\end{figure}

Again, you see that the game has
a unique Nash equilibrium, in which both firms choose the position 0.5, the mean position. We can make an argument that (0.5, 0.5) is the unique Nash equilibrium of the game as shown in Figure \ref{fig:eq2}. We can also argue that, First, (0.5, 0.5) is an equilibrium: it results in a tie, and if either firm chooses a position different from 0.5 then, it loses. Second, no other pair of positions is a Nash equilibrium, by the following
argument. (1) If one firm loses then it can do better by moving to 0.5, where it either wins outright (if other firm's position is different from 0.5) or ties for first place (if other firm's position is 0.5). (2) If the firms' tie (because their positions are symmetric about 0.5), then either firm can do better by moving to 0.5, where it wins outright.

\section{Simulation and Results}
\subsection{Data Sets Used}
\label{sec:datasets}
We used Facebook data set and Wiki-Vote data set, look their set statistics in Table \ref{fb} and Table \ref{wiki}. Facebook dataset consists of 'circles' (or 'Friends lists') from Facebook. It is a simple (undirected and unweighted) graph, people are nodes and their friendship is edge. Wiki-Vote network contains all the Wikipedia voting data from the inception of Wikipedia till January 2008. Nodes in the network represent Wikipedia users and a directed edge from node $i$ to node $j$ represents that user $i$ voted on user $j$.

\begin{table}[h!]
\caption{Facebook Dataset Statistics \cite{fbdata}}
\label{fb}
\centering
\begin{tabular}{|c|c|}
\hline
$Nodes$ & 4039 \\ \hline
$Edges$ & 88234 \\ \hline
$Nodes\ in\ largest\ WCC$ & 4039(1.000) \\ \hline
$Edges\ in\ largest\ WCC$ & 88234(1.000) \\ \hline
$Nodes\ In\ largest\ SCC$ & 4039(1.000) \\ \hline
$Edges\ in\ largest\ SCC$ & 88234(1.000) \\ \hline
$Average\ clustering\ coefficient$ & 0.6055 \\ \hline
$Number\ of\ triangles$ & 1612010 \\ \hline
$Fraction\ of\ closed\ triangles$ & 0.2647 \\ \hline
$Diameter(longest\ shortest\ path)$ & 8 \\ \hline
\end{tabular}
\end{table}

\begin{table}[h!]
\caption{Wiki-Vote Dataset Statistics \cite{wikidata1, wikidata2}}
\label{wiki}
\centering
\begin{tabular}{|c|c|}
\hline
$Nodes$ & 7115 \\ \hline
$Edges$ & 103689 \\ \hline
$Nodes\ in\ largest\ WCC$ & 7066(1.000) \\ \hline
$Edges\ in\ largest\ WCC$ & 103663(1.000) \\ \hline
$Nodes\ In\ largest\ SCC$ & 1300(1.000) \\ \hline
$Edges\ in\ largest\ SCC$ & 39456(1.000) \\ \hline
$Average\ clustering\ coefficient$ & 0.1409 \\ \hline
$Number\ of\ triangles$ & 608389 \\ \hline
$Fraction\ of\ closed\ triangles$ & 0.04564 \\ \hline
$Diameter(longest\ shortest\ path)$ & 7 \\ \hline
\end{tabular}
\end{table}

Using the datasets mentioned in the section \ref{sec:datasets}, we try to do the simulation of the proposed model discussed in section \ref{sec:model}. In this section, after spreading the information in the network we try to find how much fractions of nodes are influenced by the information and how much fractions of nodes among informed actually support the information.

In the network, there are some nodes which are equally influenced by both information. These nodes support both information with equal probability. During simulation, we assigned these nodes as a supporter of one of the information randomly. This helps in finding the exact fraction of supporter for each information for a given pair of seed nodes.

\begin{figure}[h!]
\centering
\includegraphics[width=0.6\linewidth, height=1.7 in]{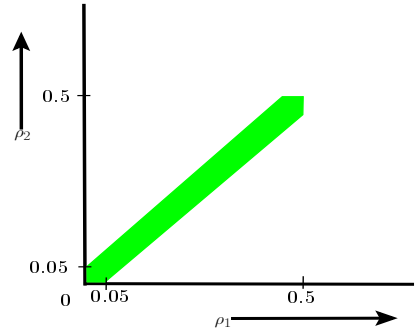}
\caption{Equilibrium region.}
\label{fig:eq_con}
\end{figure}

In order to simplify the case of a tie (or equilibrium) between the firms, a margin of 5\% is considered. Margin is nothing but the boundary for the difference between the fraction of supporter of two firms. If the difference is less than this margin, we can say equilibrium is achieved. A figure showing the equilibrium between the firms for the different fraction of supporters is shown in Figure \ref{fig:eq_con}. $\rho_1$ and $\rho_2$ are the fraction of supporter for information 1 and information 2 respectively. Green region in the Figure \ref{fig:eq_con} shows the equilibrium region.

\subsubsection{Results for Facebook and Wiki-Vote data}
\begin{figure*}[h!]
\centering
\subfigure[Fraction of influenced nodes.]{
   \includegraphics[width=0.4\linewidth, height=1.5 in] {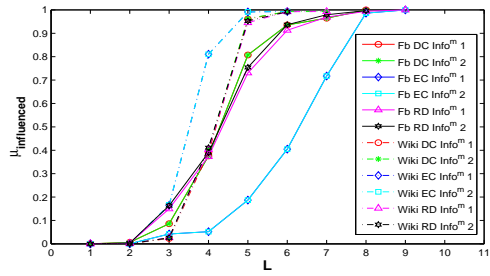}
   \label{fb_eff}
 }
 \subfigure[Fraction of supporter nodes.]{
   \includegraphics[width=0.4\linewidth, height=1.5 in]{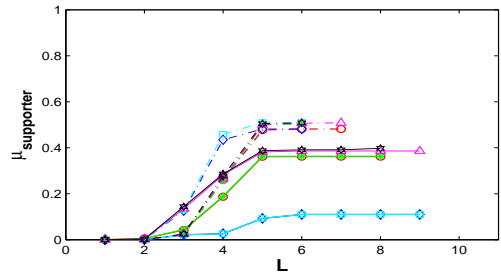}
   \label{fb_supp}
 }

\caption{Facebook and Wiki-Vote Network.}
\end{figure*}

In this section, results are shown for all three methods, DC, EC and RD on Facebook and Wiki-Vote data. As there are two competitive information (i.e., $info^m\ 1$ and $info^m\ 2$), and three methods for seed selection. Hence, for better analysis of these methods, results are further categorized into a fraction of influenced nodes ($\mu_{influenced}$) and a fraction of supporter nodes ($\mu_{supporter}$) for each information. In order to extend the analysis, random networks, tree, and random regular networks, other than Facebook dataset and Wiki-vote dataset are also generated. The properties of these networks are explained in upcoming sections. Here, $x$-axis represents number of levels `$L$' and $y$-axis represents a fraction of influenced nodes, $\mu_{influenced}$ or fraction of supporter nodes, $\mu_{supporter}$. 
In Figure \ref{fb_eff}, for Facebook network and Wiki-Vote network, how $\mu_{influenced}$ increases with \textit{L} for different methods is shown. For Facebook dataset, we can observe that $\mu_{influenced}$ for DC increase faster than EC but equivalent to RD. The main reason for this difference is the seed selection technique of these methods, which is discussed in Section \ref{intro}. Less variance is also observed for DC, EC but high variance for RD. The main reason for high variance in RD is target size. The target size is explained in Algorithm \ref{rankdegree}. The target size for seed selection is set to 10\%.
Figure \ref{fb_supp}, shows how the $\mu_{supporter}$ increases over $L$ for different methods. The main aim for any firm is to maximize their supporters. We observe that for DC, $\mu_{supporter}$ for both firms are equal at the last level. Hence, we can say equilibrium is achieved. The fraction of total population supporting one of the information is also high. Therefore, DC on Facebook network gives satisfying result. RD also gives satisfying results but with high variance. On the other hand, EC is able to achieve equilibrium but fraction of total population is less as compared to DC and RD. The reason for this behavior of EC is observed because of its seed selection technique.  The influence of information decreases highly at the starting of the diffusion process for EC.
In, Wiki-vote data, it is  observed (Figure \ref{fb_eff}) that $\mu_{influenced}$ for DC increases with levels $L$. Moderate variance is observed for DC method. Figure \ref{fb_eff} shows that for EC method, entire network is influenced by both informations. Less variance is observed for EC. RD on Wiki-vote network shows moderate variance. Similarly, for $\mu_{supporter}$ on Wiki-Vote network, DC shows (Figure \ref{fb_supp}) that significant fraction of the total population is supporting the information. As the difference between the fraction of supporter for both informations is less than the margin. Hence, equilibrium is achieved for DC but the moderate variance is also observed. Similarly, for EC and RD equilibrium is achieved and moderate variance is observed as shown in Figure \ref{fb_supp}.

Figure \ref{fb_eff} shows the combined results for all three methods DC, EC and RD for fraction of influenced nodes ($\mu_{influenced}$) on Facebbok and Wiki-Vote network. It is observed that for the Facebook dataset, DC and RD both show similar kind of behavior whereas EC behaves differently. For Wiki-vote network, DC, EC and RD all show similar kind of behavior. Finally, all three methods succeed in influencing the entire network for both datasets. Figure \ref{fb_supp} show the combined results for fraction of supporter nodes ($\mu_{supporter}$) in both networks. For the Facebook network, RD gives better results than DC and EC. Also, DC results are superior to EC. For Wiki-vote network, EC and RD give better results than DC.

\subsubsection{Results for Random Network 1}
\begin{figure*}[h!]
\centering
\subfigure[Fraction of influenced nodes.]{
   \includegraphics[width=0.4\linewidth, height=1.7 in] {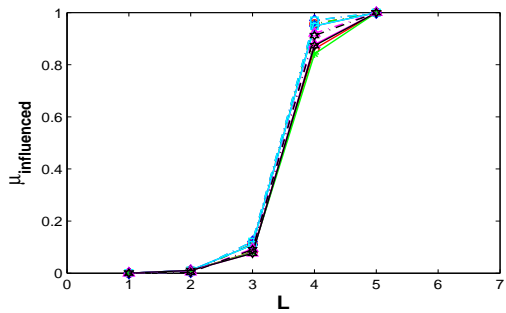}
   \label{fb_rn1_eff}
 }
 \subfigure[Fraction of supporter nodes.]{
   \includegraphics[width=0.4\linewidth, height=1.7 in]{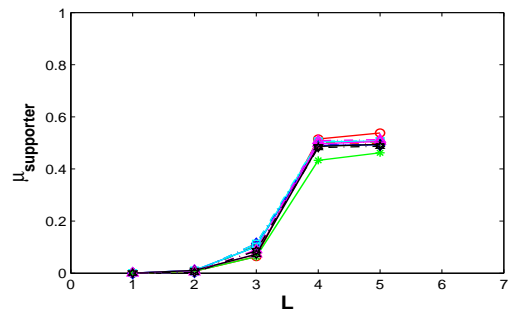}
   \label{fb_rn1_supp}
 }

\caption{Random Network 1.}
\end{figure*}

Random network 1 is a network with properties that it contains a same number of nodes as Facebook and Wiki-vote network and approximately equal average degree. The clustering coefficient (C.C.) of the random network 1 is also given for comparison with the original network. C.C. is defined as ``a measure of the degree to which nodes in a graph tend to cluster together". C.C. is a very important property for information diffusion process. It helps to understand the importance of loops and clusters in the network. 
For random network 1 with similar properties to Facebook network, Figure \ref{fb_rn1_eff} shows that in less number of levels complete network is influenced by information compared to original network. This phenomenon is observed for all methods (DC, EC and RD). We can conclude that the diameter of random network 1 is less than original network. $\mu_{supporter}$ for both information is shown in Figure \ref{fb_rn1_supp}. For DC, less variance is observed and also equilibrium is achieved. Same behavior like DC is shown by EC (Figure \ref{fb_rn1_supp}). 
In Figure \ref{fb_rn1_supp}, supporter for information 1 and information 2 are shown. For DC, it is observed that the difference between the $\mu_{supporter}$ for both information is greater than the margin (5\%). Hence, equilibrium is not achieved. For EC, there is a significant improvement in fraction of supporter for information compared to original network. The main reason for this improvement is diameter of the network. As we explained earlier, EC chooses seed node which is closer to high degree node. If the number of level increases, then the influence of information decreases drastically with each increasing level in the network. The difference between $\mu_{supporter}$ for both information is less than margin. Therefore, equilibrium is achieved for EC. For RD method also $\mu_{supporter}$ is satisfactory for both information and also equilibrium is achieved.
Similarly, random network 1 is a network with properties that it contains an equal number of nodes as Wiki-vote network and approximately equal average degree of the network. The C.C. of this network is also very less compared to Wiki-vote network. This means that there are less number of loops in the network.
Figure \ref{fb_rn1_eff} shows that for DC number of levels for influence of information is five. Similarly, for EC and RD number of levels are five. Therefore, we can conclude that diameter for random network 1 is less compared to original Wiki-Vote network.
In Figure \ref{fb_rn1_supp} supporters for both informations are shown. For DC, equilibrium is achieved but moderate variance is observed. Similarly, for EC and RD equilibrium is achieved and moderate variance is observed.

For all three methods DC, EC and RD for fraction of influenced nodes ($\mu_{influenced}$) in the random network 1. It is observed that, DC, EC and RD all show similar kind of behavior. Finally, all three methods succeed in influencing the entire network. For fraction of supporter nodes ($\mu_{supporter}$) RD and EC give better results than DC on random network 1 with similar properties to Facebook network. For random network 1 with similar properties to Wiki-Vote data, all methods show similar results.

\subsubsection{Results for Random Network 2}
\begin{figure*}[h!]
\centering
\subfigure[Fraction of influenced nodes.]{
   \includegraphics[width=0.4\linewidth, height=1.7 in] {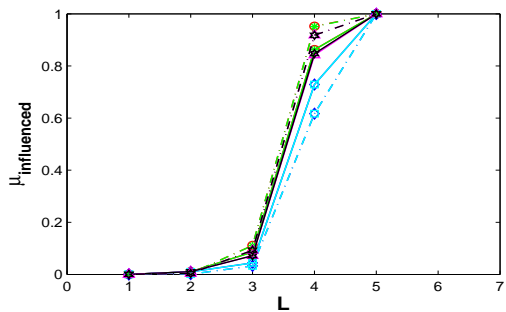}
   \label{fb_rn2_eff}
 }
 \subfigure[Fraction of supporter nodes.]{
   \includegraphics[width=0.4\linewidth, height=1.7 in]{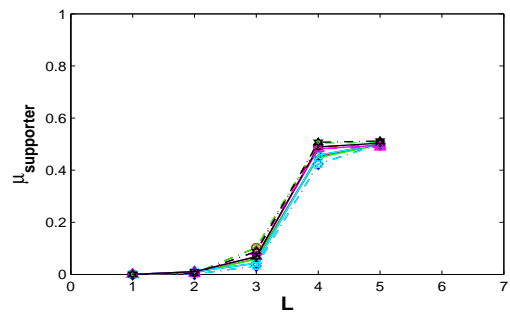}
   \label{fb_rn2_supp}
 }

\caption{Random Network 2.}
\end{figure*}

Random network 2 also contains the same properties as random network 1. For random network 2 with similar properties to Facebook network, Figure \ref{fb_rn2_eff} shows that for DC the entire network is influenced by information in four levels. For EC it takes five levels and for RD, the process takes four levels. Hence, we can conclude that the diameter of the network is less compared to Facebook network.
$\mu_{supporter}$ for both informations are shown in Figure \ref{fb_rn2_supp}. For DC, less variance is observed and also equilibrium is achieved. Same behavior like DC is shown by EC. For RD, moderate variance is observed but equilibrium is achieved.

For random network 2 with similar properties to Wiki-Vote network, Figure \ref{fb_rn2_eff}, shows the combined results for all three methods DC, EC and RD for fraction of influenced nodes ($\mu_{influenced}$). It is observed that for random network 2, DC, EC and RD all shows similar kind of behavior. Finally, all three methods succeed in influencing the entire network. Figure \ref{fb_rn2_supp}, shows the combined results for fraction of supporter nodes ($\mu_{supporter}$) in the network. For random network 2, DC, EC and RD all gives same results.

\begin{figure*}[h!]
\centering
\subfigure[Fraction of influenced nodes.]{
   \includegraphics[width=0.4\linewidth, height=1.7 in] {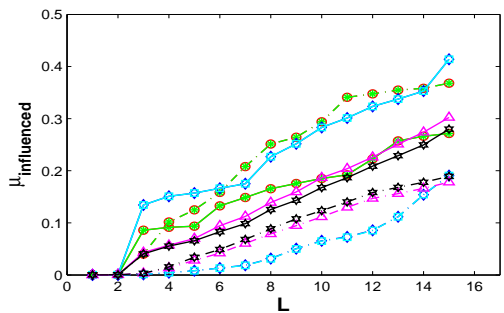}
   \label{fb_st_eff}
 }
 \subfigure[Fraction of supporter nodes.]{
   \includegraphics[width=0.4\linewidth, height=1.7 in]{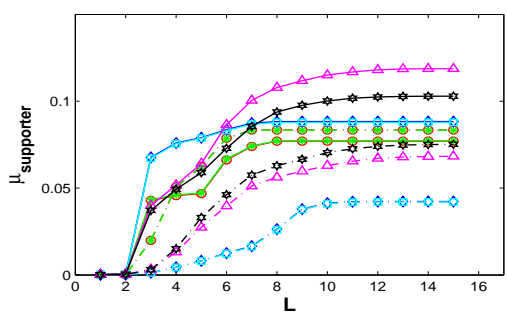}
   \label{fb_st_supp}
 }

\caption{Tree Network.}
\end{figure*}

\begin{figure*}[h!]
\centering
\subfigure[Fraction of influenced nodes.]{
   \includegraphics[width=0.4\linewidth, height=1.7 in] {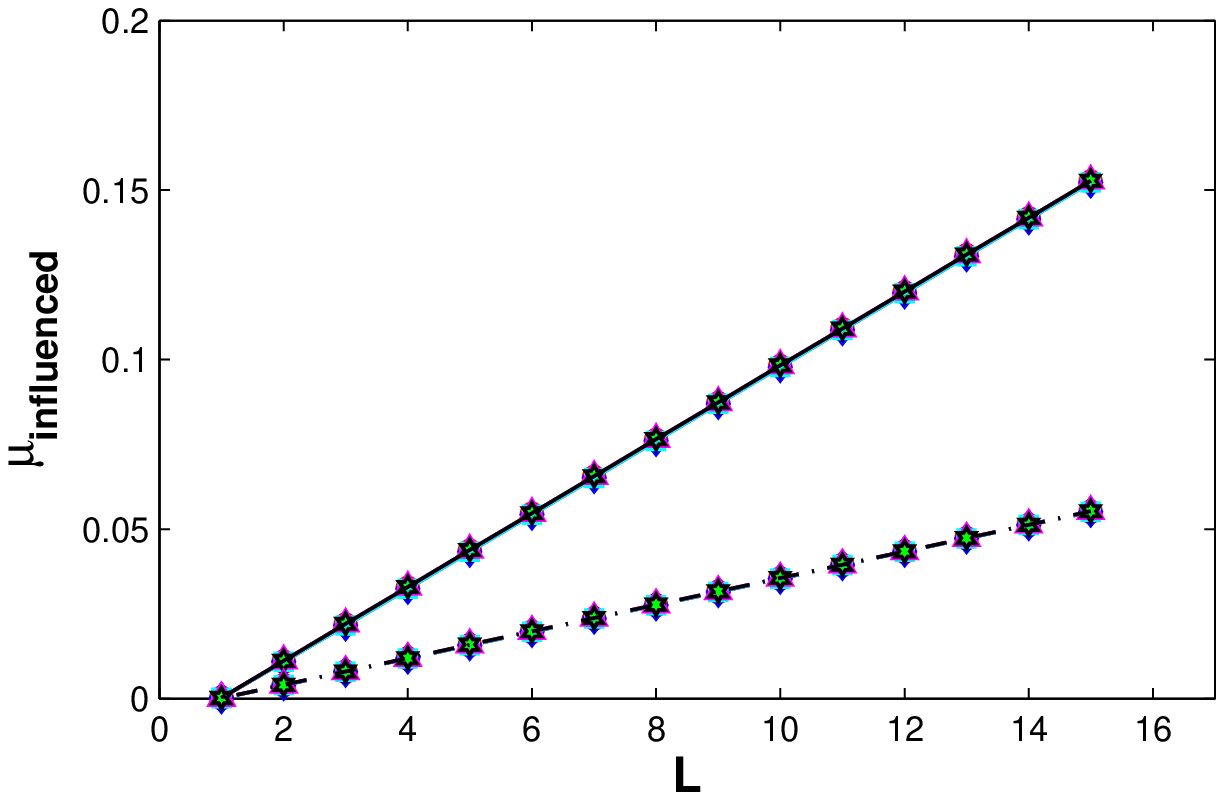}
   \label{fb_reg_eff}
 }
 \subfigure[Fraction of supporter nodes.]{
   \includegraphics[width=0.4\linewidth, height=1.7 in]{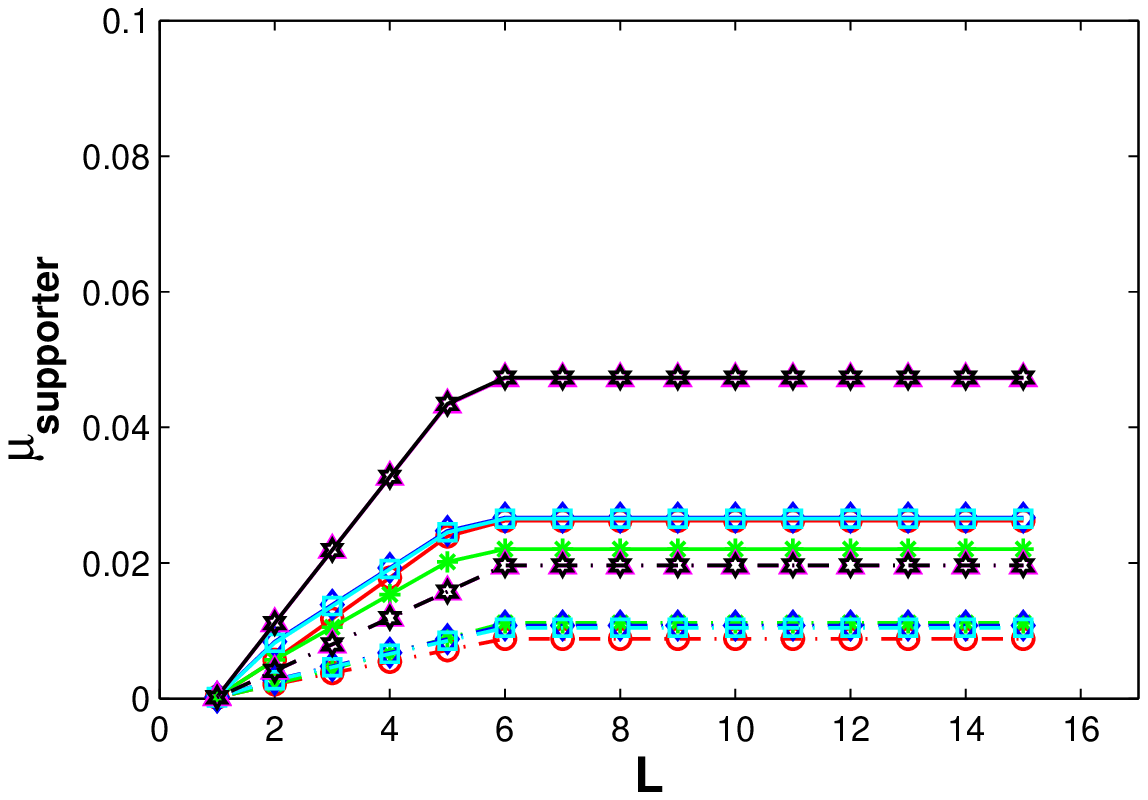}
   \label{fb_reg_supp}
 }

\caption{Regular Network.}
\end{figure*}
\subsubsection{Results for Random Tree}
Random tree network is a tree generated by converting original network (Facebook and Wiki-Vote) into a spanning tree. The Clustering Coefficient of the random tree is zero. Hence, it ensures that there is no loop in the tree network. As we discussed earlier, the loop is very important for the information diffusion process. Now, we can show this experimentally by comparing the random tree network results with the original networks.
For spanning tree from Facebook network, Figure \ref{fb_st_eff} shows that for DC less fraction of nodes of the total population is influenced by the information. Similar behavior is observed by RD. On the other hand, EC is able to influence larger population compared to DC and RD. This kind of behavior is observed for DC, EC, and RD because information is not able to pass the seed node of other information. It is discussed in section \ref{InformationSpreading} using Figure \ref{fig:icm1}. Very less variance is observed for DC and EC but moderate variance for RD.
As there is no loop or cluster, influence of information diminishes very quickly at each level. For DC, as shown in Figure \ref{fb_st_supp}, $\mu_{supporter}$ is less but equilibrium is achieved. Similarly, EC $\mu_{supporter}$ is less but equilibrium is achieved. For RD, $\mu_{supporter}$ is less and equilibrium is achieved but high variance is observed (Figure \ref{fb_st_supp}).

For spanning tree from Wiki-Vote network, Figure \ref{fb_st_eff} shows that for DC less fraction of nodes of the total population is influenced by the information. Similar behavior is observed by RD. On the other hand, EC is able is to influence larger population compared to DC and RD (Figure \ref{fb_st_eff}). Very less variance is observed for DC and EC but moderate variance for RD.
For DC, as shown in Figure \ref{fb_st_supp}, $\mu_{supporter}$ is less but equilibrium is achieved. Similarly, EC $\mu_{supporter}$ is less but equilibrium is achieved (Figure \ref{fb_st_supp}). For RD, $\mu_{supporter}$ is less and equilibrium is achieved but the high variance is observed (Figure \ref{fb_st_supp}).
It is observed that for random tree network from Wiki-Vote network,DC gives better results than EC and RD. Also, RD gives better results than EC.

\subsubsection{Results for Regular Network}
The regular network is a random network with approximately equal average degree of the original network (Facebook and Wiki-Vote). The degree distribution of nodes in a network also plays an important role in the information diffusion process. The random regular network helps us to understand the importance of degree distribution of nodes in a network.
For regular network with similar properties to Facebook network, Figure \ref{fb_reg_eff}, shows the fraction of influenced nodes. For DC, less fraction of nodes of total population is influenced by information. Similar behavior is observed for EC and RD (Figure \ref{fb_reg_eff}). Similarly, Figure \ref{fb_reg_supp} shows the fraction of supporter nodes. For DC, very less $\mu_{supporter}$ and moderate variance is observed but equilibrium is achieved. For EC and RD, very less $\mu_{supporter}$ and less variance but equilibrium is achieved (\ref{fb_reg_supp}). Overall, RD gives much better results than DC and EC.

For regular network with similar properties to Wiki network, Figure \ref{fb_reg_eff} shows that for DC less fraction of nodes of total population is influenced by information. Similar behavior is observed for EC and RD.
For DC very less $\mu_{supporter}$ and moderate variance is observed but equilibrium is achieved. For EC and RD, very less $\mu_{supporter}$ and less variance but equilibrium is achieved (Figure \ref{fb_reg_supp}). Finally, RD gives much better results than DC and EC.


These results help to understand
\begin{enumerate}
\item \textbf{Importance of topological loops.}
Network structure is important in spreading information. Topological loop is one of the network property. The importance of topological loops is shown by comparing Facebook network results with random tree. It is observed that spreading slows down in the tree network. Fraction of supporter decreases drastically.

\item \textbf{Importance of degree distribution} can be understood by comparing results of random regular networks with Facebook networks. The fraction of supporters in random regular network is less. Hence, network distribution is very important for spreading process. Similarly, importance of network properties can be seen for Wiki-vote dataset.
\end{enumerate}

\begin{figure}[h!]
\centering
\includegraphics[width=1\linewidth, height=1.3 in]{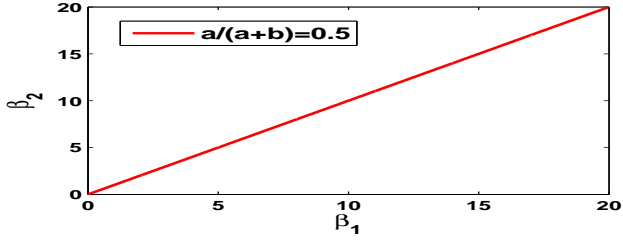}
\caption{For $a/(a+b)$ = 0.5}
\label{fig:1a}
\end{figure}

Plot for values of ($\beta_1,\beta_2$), when [$a/(a+b)$]=0.5 is shown in Figure \ref{fig:1a}. This will unfold the equilibrium state points values for $\beta_1$ and $\beta_2$. 

\begin{figure}[h!]
\centering
\includegraphics[width=1\linewidth, height=1.3 in]{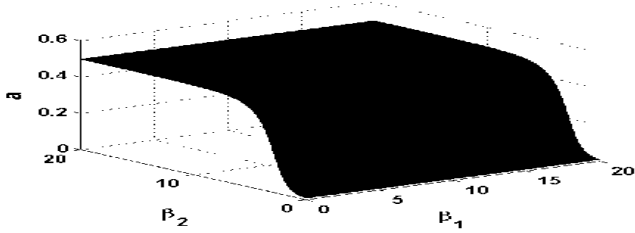}
\caption{Surface plot for $a$, where $a/(a+b)$=0.5}
\label{fig:1b}
\end{figure}

Surface plot for $a$ on different values of ($\beta_1,\beta_2$) when [$a/(a+b)$]=0.5 is shown in Figure \ref{fig:1b}. These plots unfolds the values of $a$ at equilibrium against various values of $\beta_1$ and $\beta_2$.

\begin{figure}[h!]
\centering
\includegraphics[width=1\linewidth, height=1.3 in]{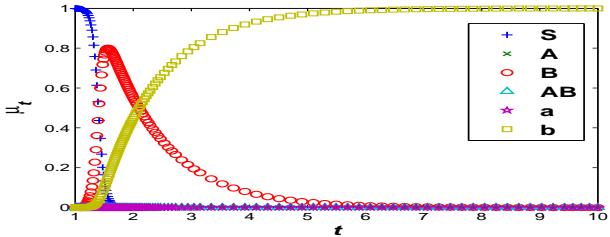}
\caption{Time evolution of supporter nodes : For $\beta_1$=1 and $\beta_2$=20, $a$ is very low but $b$ is very high.}
\label{fig:timeplot1}
\end{figure}

\begin{figure}[h!]
\centering
\includegraphics[width=1\linewidth, height=1.3 in]{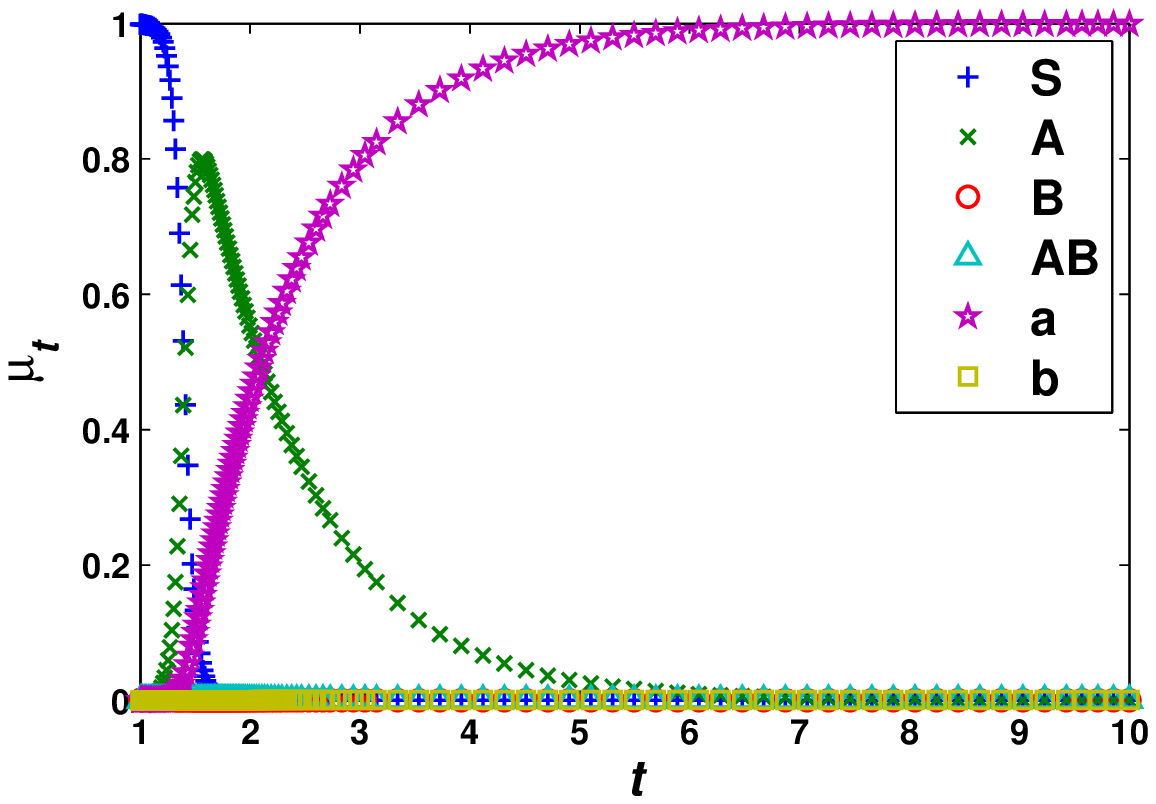}
\caption{Time evolution of supporter nodes : For $\beta_1$=20 and $\beta_2$=1, $b$ is very low but $a$ is very high.}
\label{fig:timeplot2}
\end{figure}

\begin{figure}[h!]
\centering
\includegraphics[width=1\linewidth, height=1.3 in]{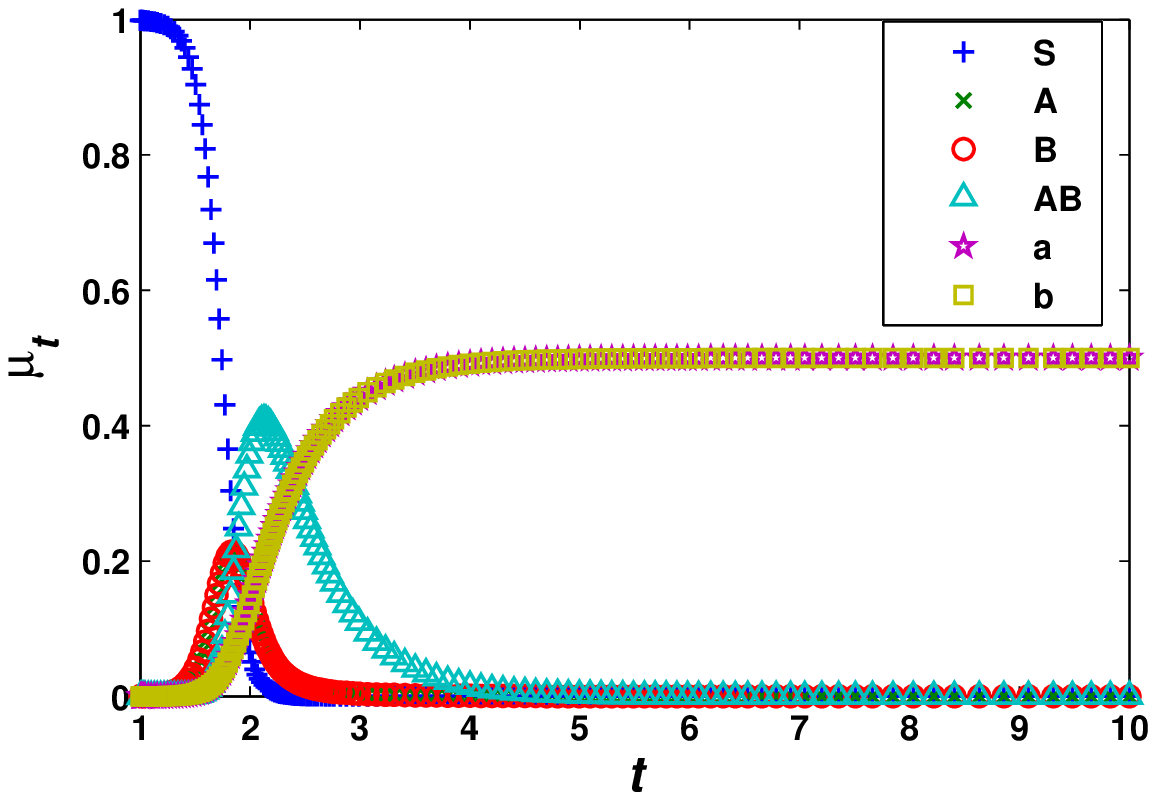}
\caption{Time evolution of supporter nodes : For $\beta_1$=10 and $\beta_2$=10, $a$ is very low but $b$ is very high.}
\label{fig:timeplot3}
\end{figure}

\begin{figure}[h!]
\centering
\includegraphics[width=1\linewidth, height=1.3 in]{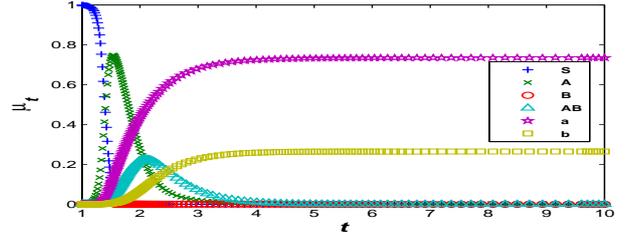}
\caption{Time evolution of supporter nodes : For $\beta_1$=20 and $\beta_2$=10, $b$ is very low but $a$ is very high.}
\label{fig:timeplot4}
\end{figure}

For analyzing the change in various compartments with time, fraction of nodes (i.e., $\mu$) against time are shown in Figure \ref{fig:timeplot1} - \ref{fig:timeplot4}. Initial conditions are: \textbf{S(0)}=0.999, \textbf{A(0)}=0.0005, \textbf{B(0)}=0.0005, \textbf{AB(0)}=0, \textit{a(0)}=0 and \textit{b(0)}=0. Hence, \textit{Time Interval} denotes how values for variable changes in differential equations for input between 1 to 10. In order to understand the effect of $\beta_1$ and $\beta_2$ on compartments, four different cases are considered: (1) $\beta_1$=1 and $\beta_2$=20, (2) $\beta_1$=20 and $\beta_2$=1, (3) $\beta_1$=10 and $\beta_2$=10, and (4) $\beta_1$=20 and $\beta_2$=10.

\begin{figure}[h!]
        \centering
        \includegraphics[width=0.7\linewidth, height=1.7 in]{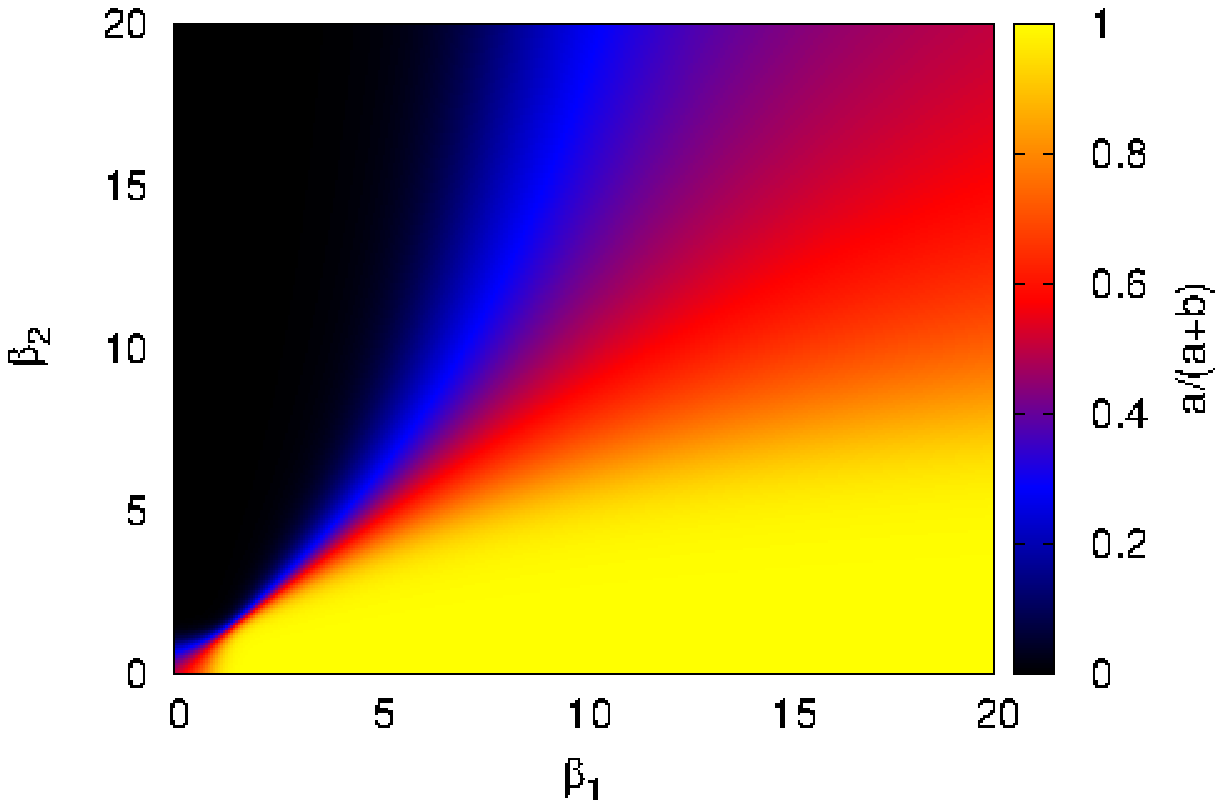}
        \caption{Plot for $a/(a+b)$, when \textbf{A},\textbf{B} and \textbf{AB} =0}
        \label{fig:3a}
\end{figure}

Steady state of any dynamics is achieved only when all their nodes in active state(s), i.e. \textbf{A}, \textbf{B} and \textbf{AB}, become zero. Steady state plot is shown in Figure \ref{fig:3a}. It shows that for different combinations of $\beta_1$ and $\beta_2$ ( where, 0$\leq\beta_1$,$\beta_2\leq$20 ), how much fraction of total supporter lies in compartment $a$ and $b$. 

\section{Conclusions}
\textbf{Cascade model}: There are two parameters on which we can compare results (1) Clustering Coefficient, and (2) Degree distribution of nodes.

In order to compare methods, all these methods DC, EC, and RD are very different from each other. Here, the main difference in results is observed due to seed node selection method. And this seed node selection directly depends upon network structure. Hence, the difference we observed in results is due to (1) seed selection method, and (2) network structure parameters i.e., clustering coefficient and degree distribution of nodes.

By comparing the results of trees and Facebook network we can say that the topological loops are important for dynamics and as a consequence, seed selection algorithms not work well. And also results compared to random and random regular networks say how much the degree distributions are important for dynamics.

On comparing results on both datasets, (1) Degree centrality gives better results on both datasets as total fraction of nodes support any information is high. And also a fraction of nodes support each information are in equilibrium. DC shows this kind of behavior because it gives importance to node degree for seed selection. (2) Eigenvector centrality shows the better results for Wiki-vote dataset but not for the Facebook dataset. This behavior is observed because eigenvector centrality gives importance to the node whose neighbors are highly connected. In the Facebook dataset, the average degree of a nodes is very high and hence, the degree of maximum degree node is also high. Therefore, the influence of information diminishes very fast in the Facebook dataset. But in the Wiki-vote dataset, the average degree is less, so, the influence of information takes the time to reach its influence less than the threshold. (3) Rank degree method considers the sampling of the network for seed selection. This helps it in choosing a node from which target size of the network is easily reachable. This method works equally good on both network and also better than other two methods. This behavior is observed because of its seed selection technique.
\\
\textbf{Mean field model}: From the experimental results, we can conclude that:
\begin{enumerate}
\item If fraction of initial seed nodes for firms and rate of information spreading are less, then most of the nodes in the network remains in the uninformed compartment, \textbf{S}. As we can see from figure \ref{fig:SS-inf}, fraction of initial seed nodes and rate of information spreading are inversely proportonal to each other in order to inform large fraction of nodes.
\begin{figure}[h!]
\centering
\includegraphics[width=0.7\linewidth, height=1.7 in]{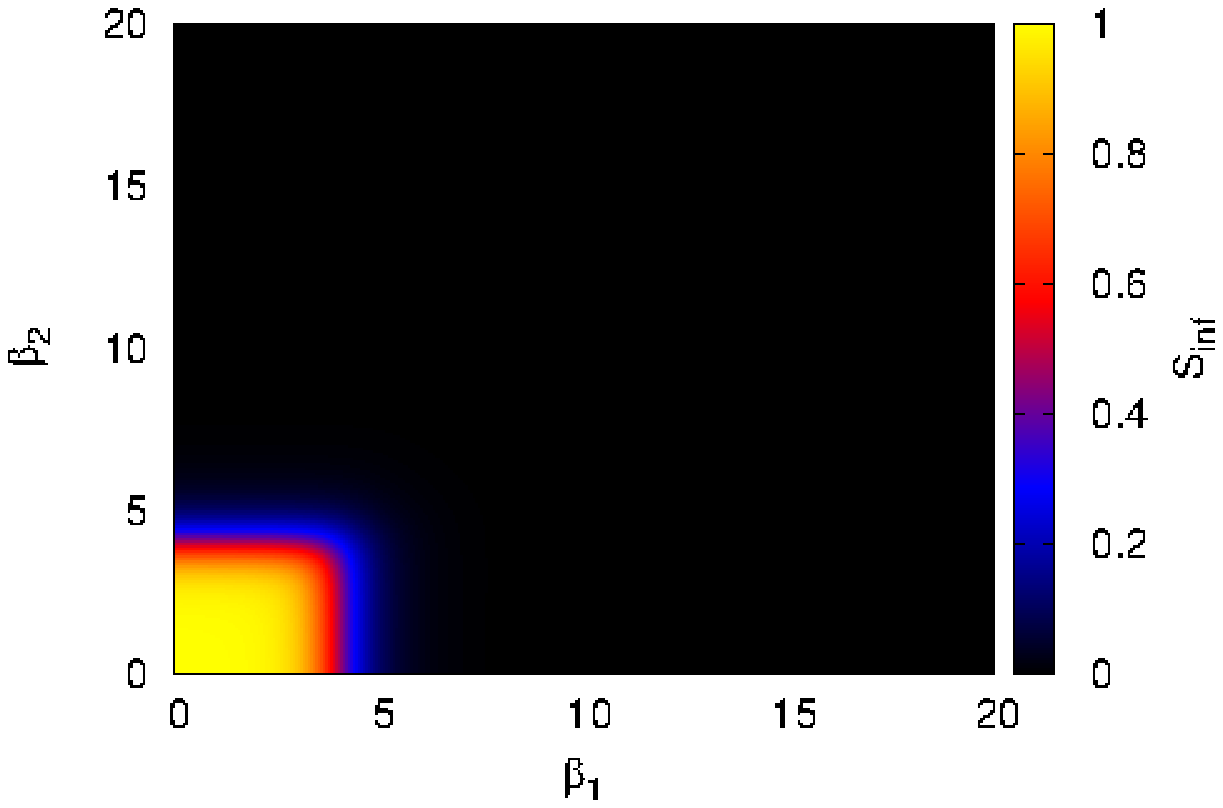}
\caption{Mean Field Model : A=0.0005, B=0.0005}
\label{fig:SS-inf}
\end{figure}
\begin{figure}[h!]
\centering
\includegraphics[width=0.7\linewidth, height=1.7 in]{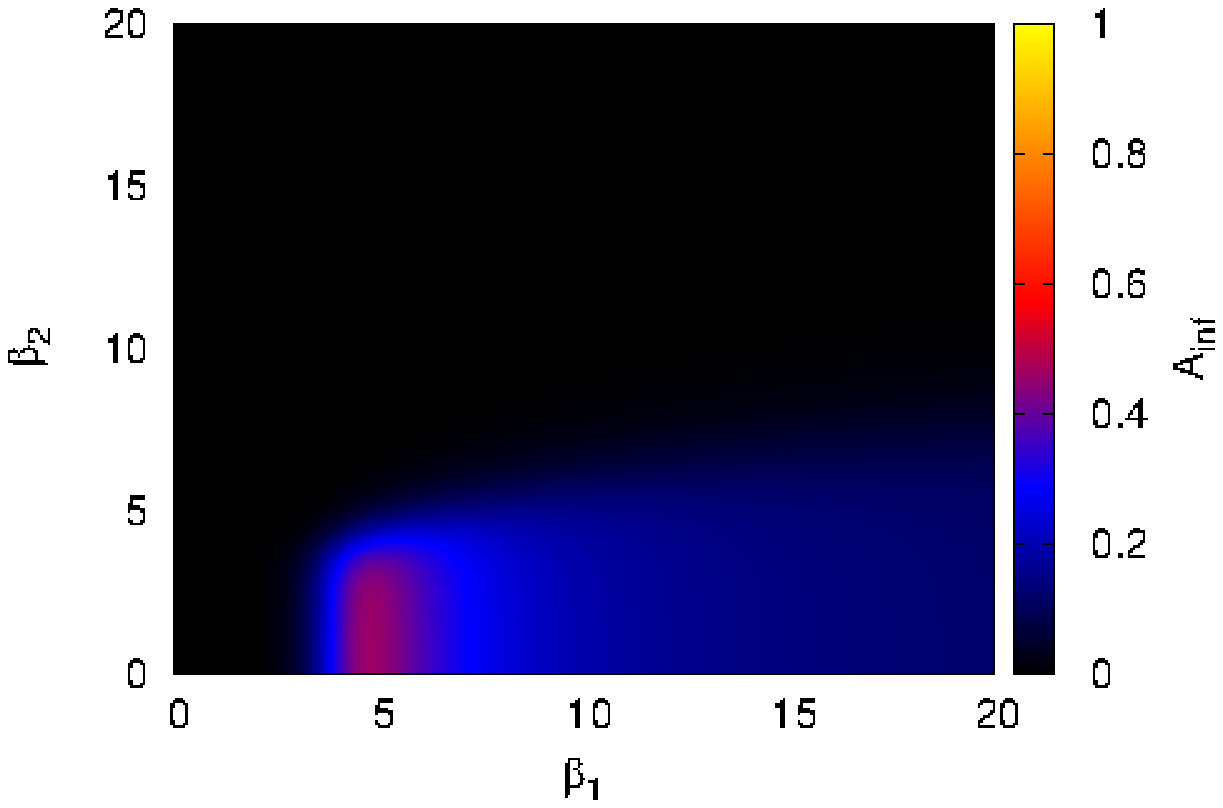}
\caption{Mean Field Model : A=0.0005, B=0.0005}
\label{fig:AA-inf}
\end{figure}
\item From Figure \ref{fig:AA-inf}, we can conclude that at some specific value of $\beta_1$, maximum fraction of informed nodes by information 1 is observed. If we increase this value of $\beta_1$, then informed nodes for information 1 decreases. Similar behavior is observed for $\beta_2$.
\item Fraction of nodes informed by one or both information and supporter nodes for different information, directly depends upon fraction of initial nodes and information spreading rate.
\end{enumerate}



\clearpage
\appendix
\onecolumn
\section{Numerical solution for the Proposed Mean Field Model} \label{mf}
\label{app}

Numerical simulations of the proposed meal field model to understand more clearly, we took different fraction of seed nodes for spreading information. Comparison is shown in Figures A.25-A.30, for three different initial values of \textbf{A} and \textbf{B}, i.e., (a) \textbf{A}=0.0005, \textbf{B}=0.0005, (b) \textbf{A}=0.001, \textbf{B}=0.001 and (c) \textbf{A}=0.01, \textbf{B}=0.01.

\begin{figure*}[ht!]
\centering
\subfigure[A=0.0005, B=0.0005]{
   \includegraphics[width=0.3\linewidth, height=1.8 in] {Figures/S_ode.eps}
   \label{fig:S-inf}
 }
\subfigure[A=0.001, B=0.001]{
   \includegraphics[width=0.3\linewidth, height=1.8 in] {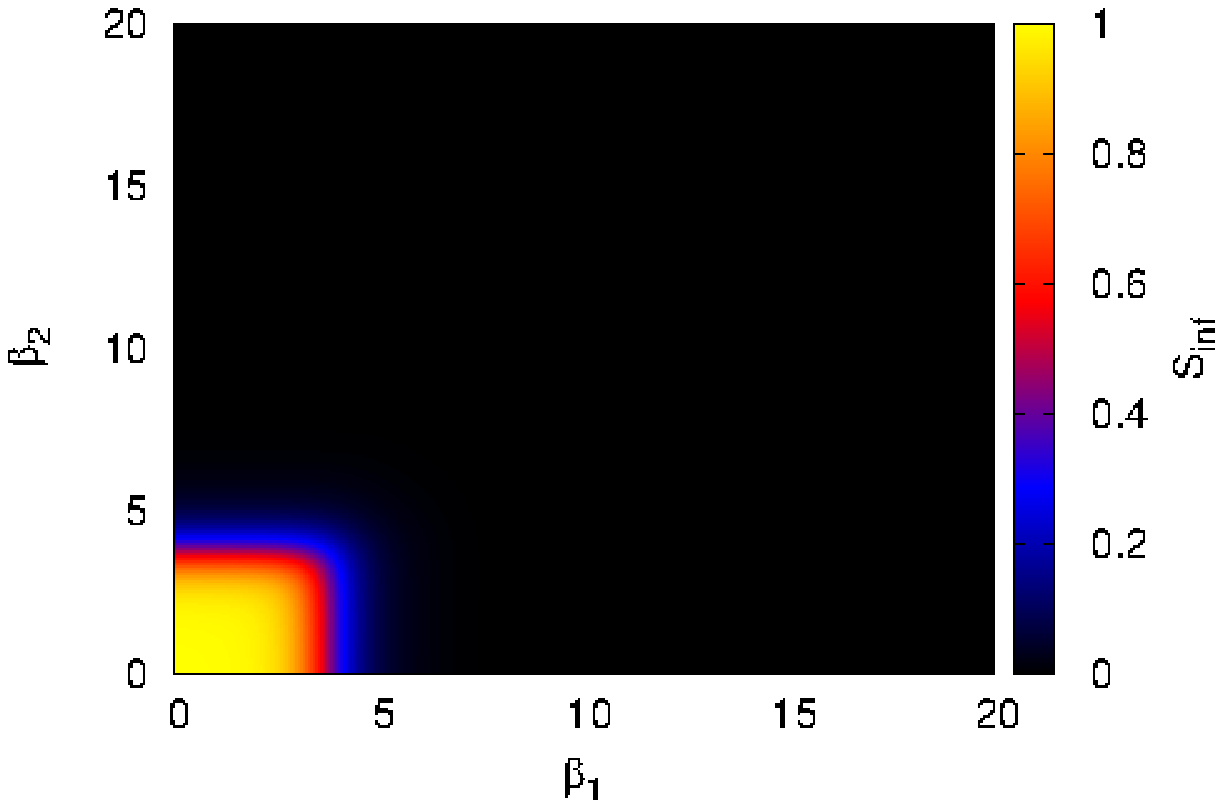}
   \label{fig:S1-inf}
 }
 \subfigure[A=0.01, B=0.01]{
   \includegraphics[width=0.3\linewidth, height=1.8 in]{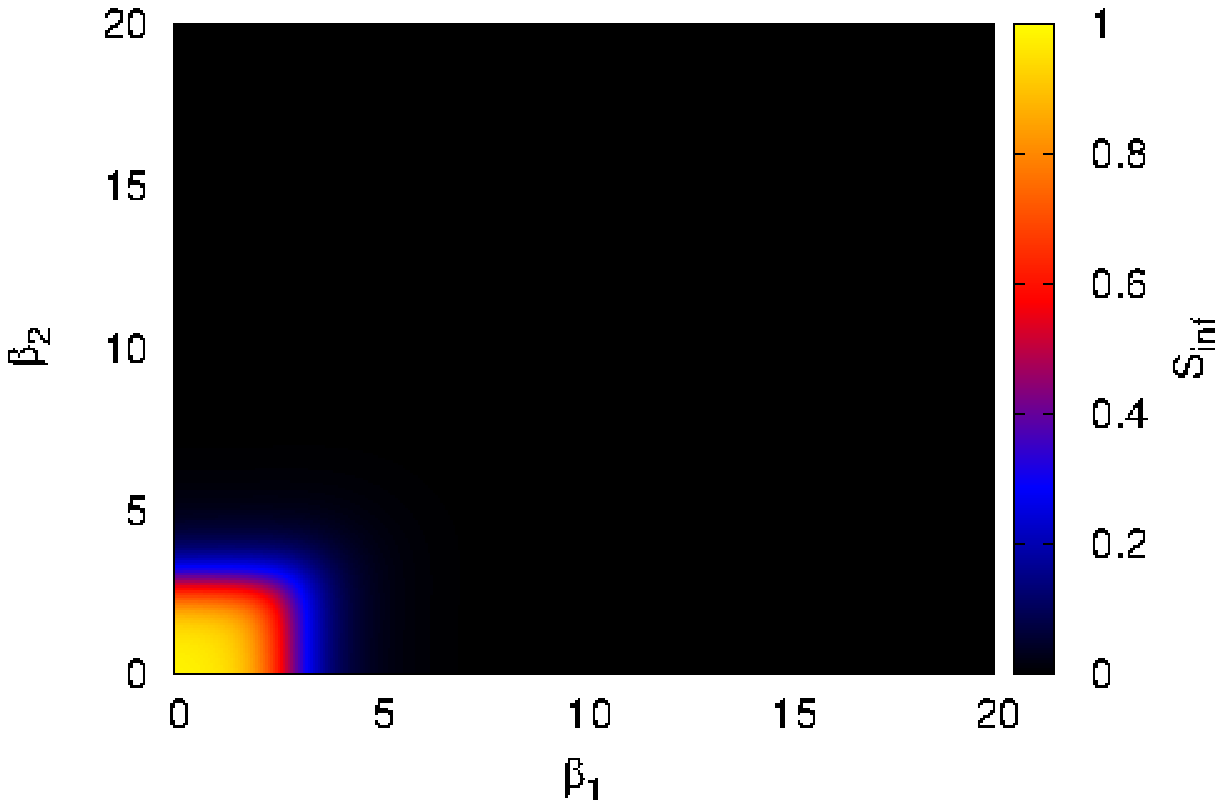}
   \label{fig:S2-inf}
 }
\label{fig:S-infinity}
\caption{This shows how fraction of uninformed nodes changes with values of $\beta_1$ and $\beta_2$ as we change initial values of $A$ and $B$. As result shows on increasing initial values of $A$ and $B$, $S$ will diminishes quickly.}
\end{figure*}

\begin{figure*}[ht!]
\centering
\subfigure[A=0.0005, B=0.0005]{
   \includegraphics[width=0.3\linewidth, height=2 in] {Figures/A_ode.eps}
   \label{fig:A-inf}
 }
\subfigure[A=0.001, B=0.001]{
   \includegraphics[width=0.3\linewidth, height=2 in] {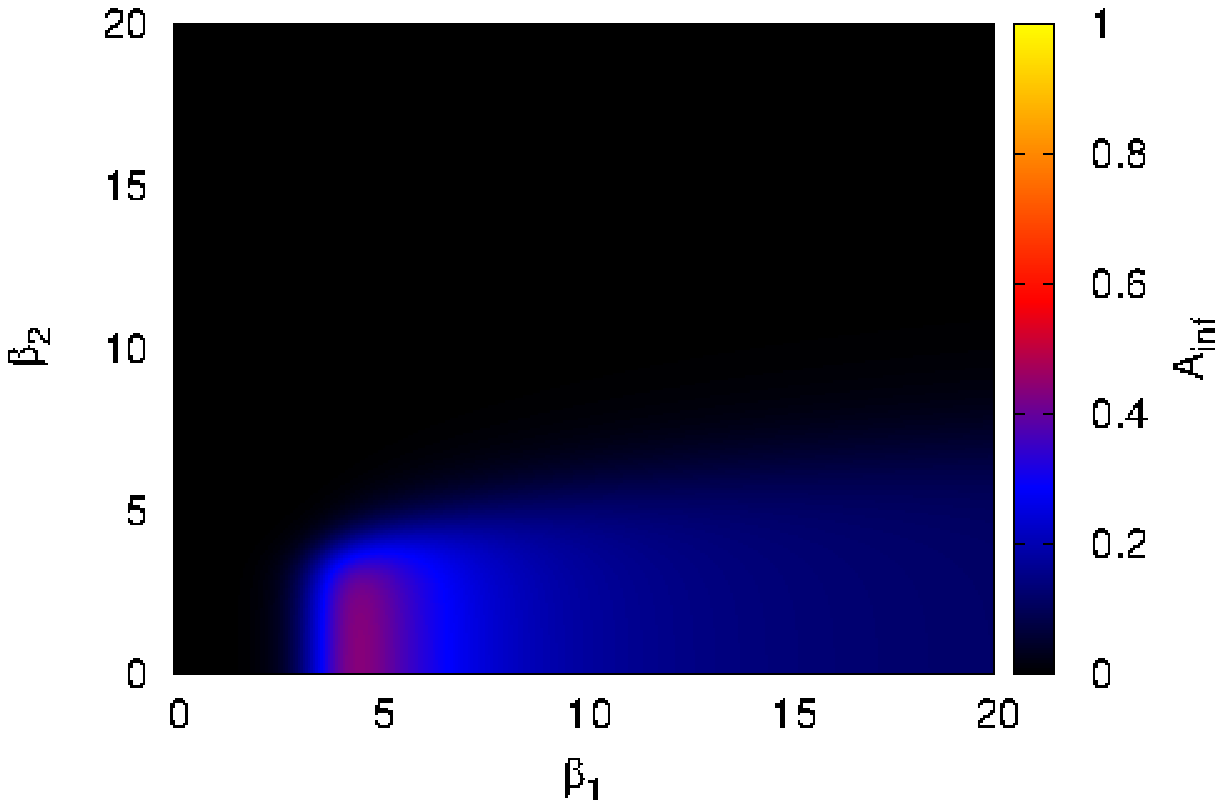}
   \label{fig:A1-inf}
 }
 \subfigure[A=0.01, B=0.01]{
   \includegraphics[width=0.3\linewidth, height=2 in]{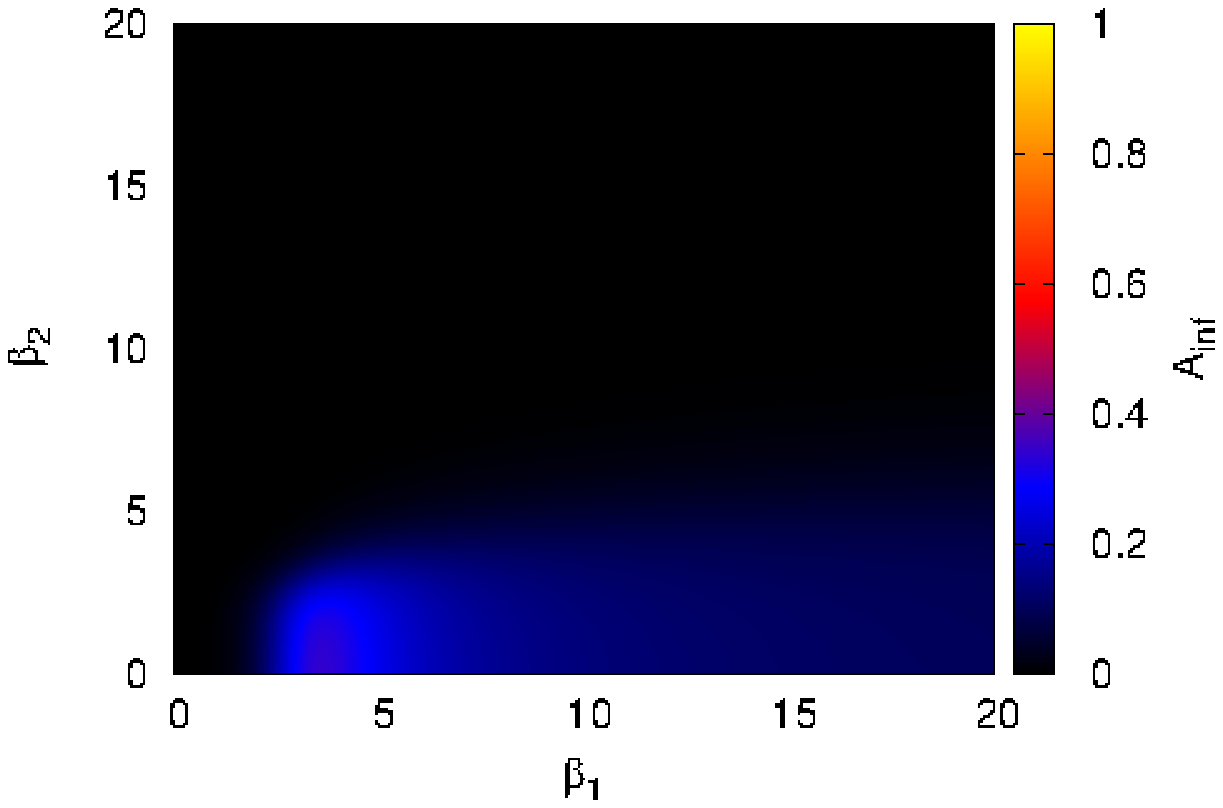}
   \label{fig:A2-inf}
 }
\label{A_infinity}
\caption{This shows how fraction of informed nodes with information 1 changes with values of $\beta_1$ and $\beta_2$ as we change initial values of $A$ and $B$. As result shows on increasing initial values of $A$ and $B$, fraction of nodes informed by information 1 will reach their peak value for less value of $\beta_1$.}
\end{figure*}

\begin{figure*}[ht!]
\centering
\subfigure[A=0.0005, B=0.0005]{
   \includegraphics[width=0.3\linewidth, height=2 in] {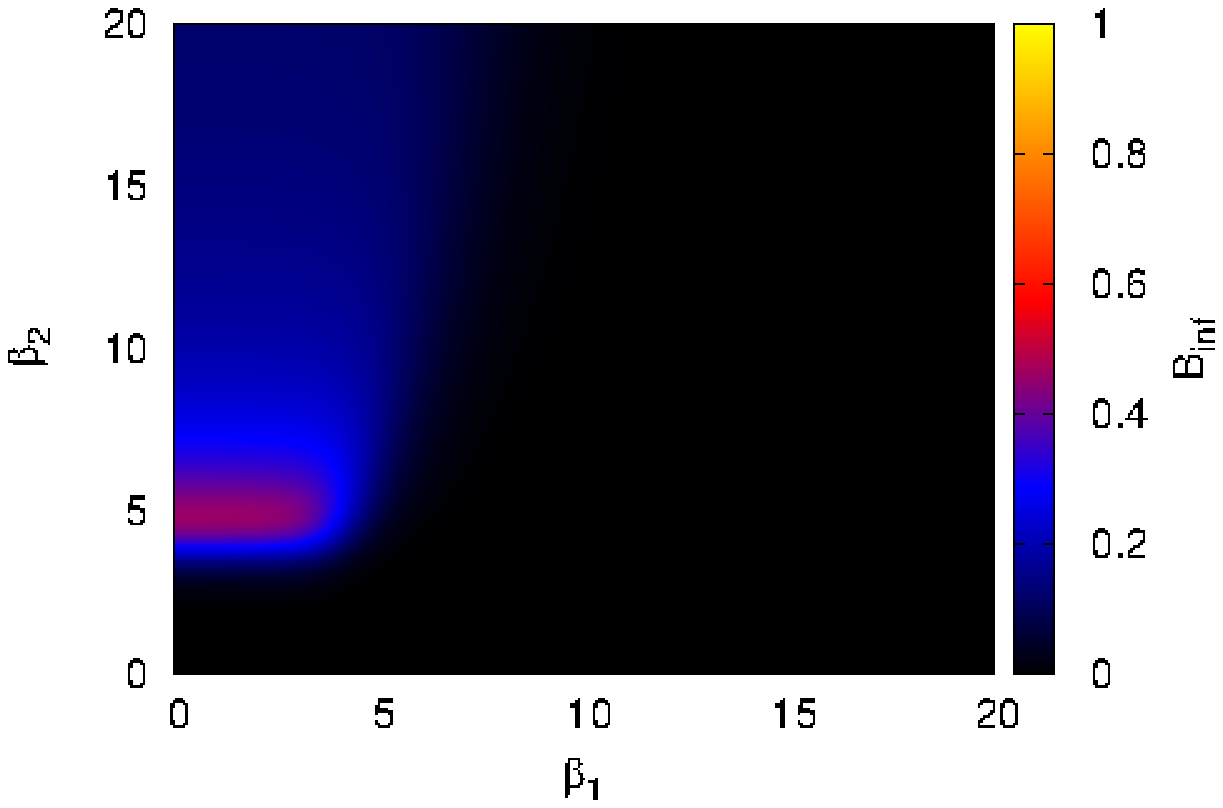}
   \label{fig:B-inf}
 }
\subfigure[A=0.001, B=0.001]{
   \includegraphics[width=0.3\linewidth, height=2 in] {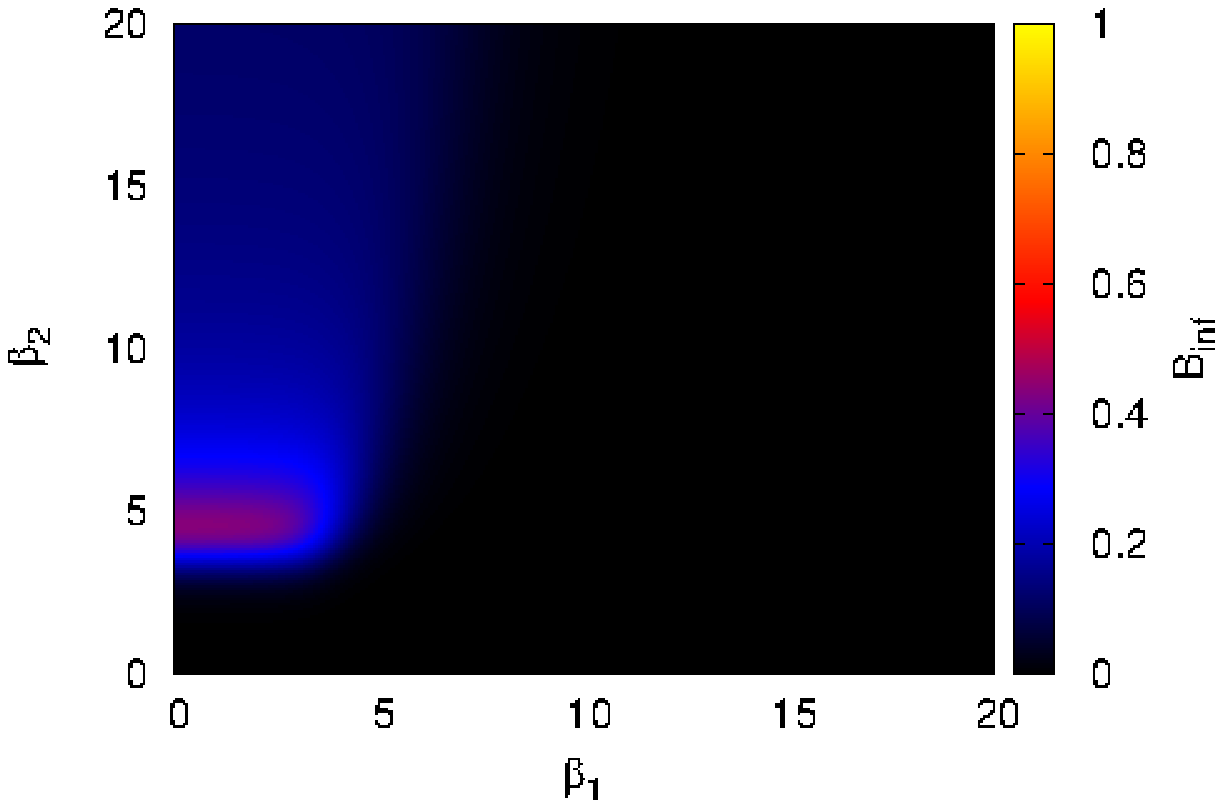}
   \label{fig:B1-inf}
 }
 \subfigure[A=0.01, B=0.01]{
   \includegraphics[width=0.3\linewidth, height=2 in]{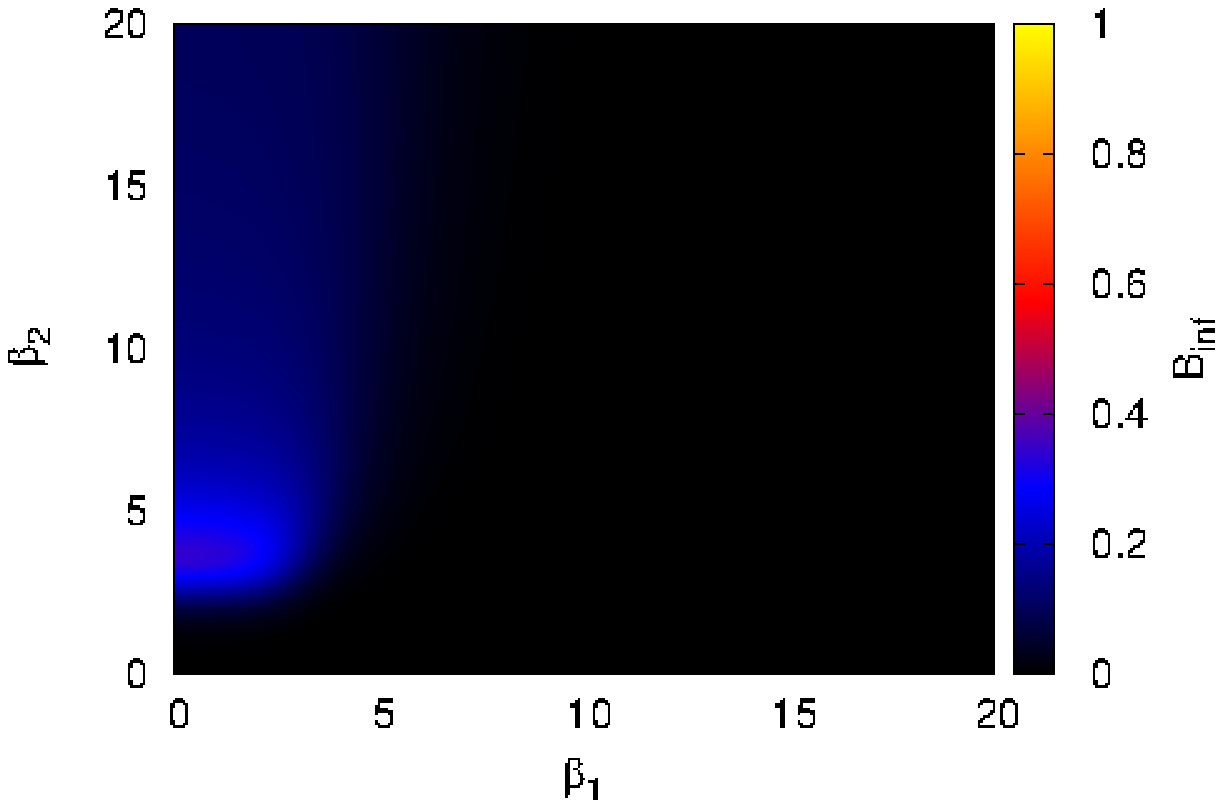}
   \label{fig:B2-inf}
 }
\label{B_infinity}
\caption{This shows how fraction of informed nodes with information 2 changes with values of $\beta_1$ and $\beta_2$ as we change initial values of $A$ and $B$. As result shows on increasing initial values of $A$ and $B$, fraction of nodes informed by information 2 will reach their peak value for less value of $\beta_2$.}
\end{figure*}

\begin{figure*}[ht!]
\centering
\subfigure[A=0.0005, B=0.0005]{
   \includegraphics[width=0.3\linewidth, height=2 in] {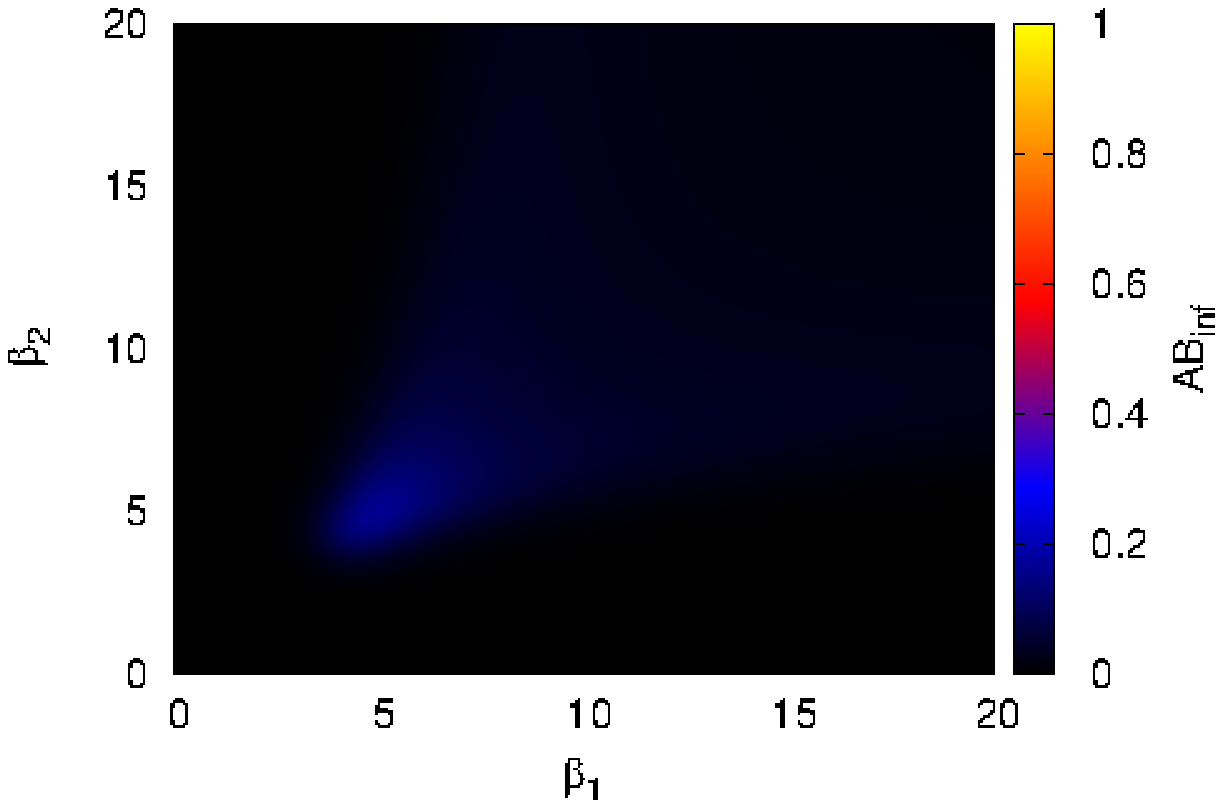}
   \label{fig:AB-inf}
 }
\subfigure[A=0.001, B=0.001]{
   \includegraphics[width=0.3\linewidth, height=2 in] {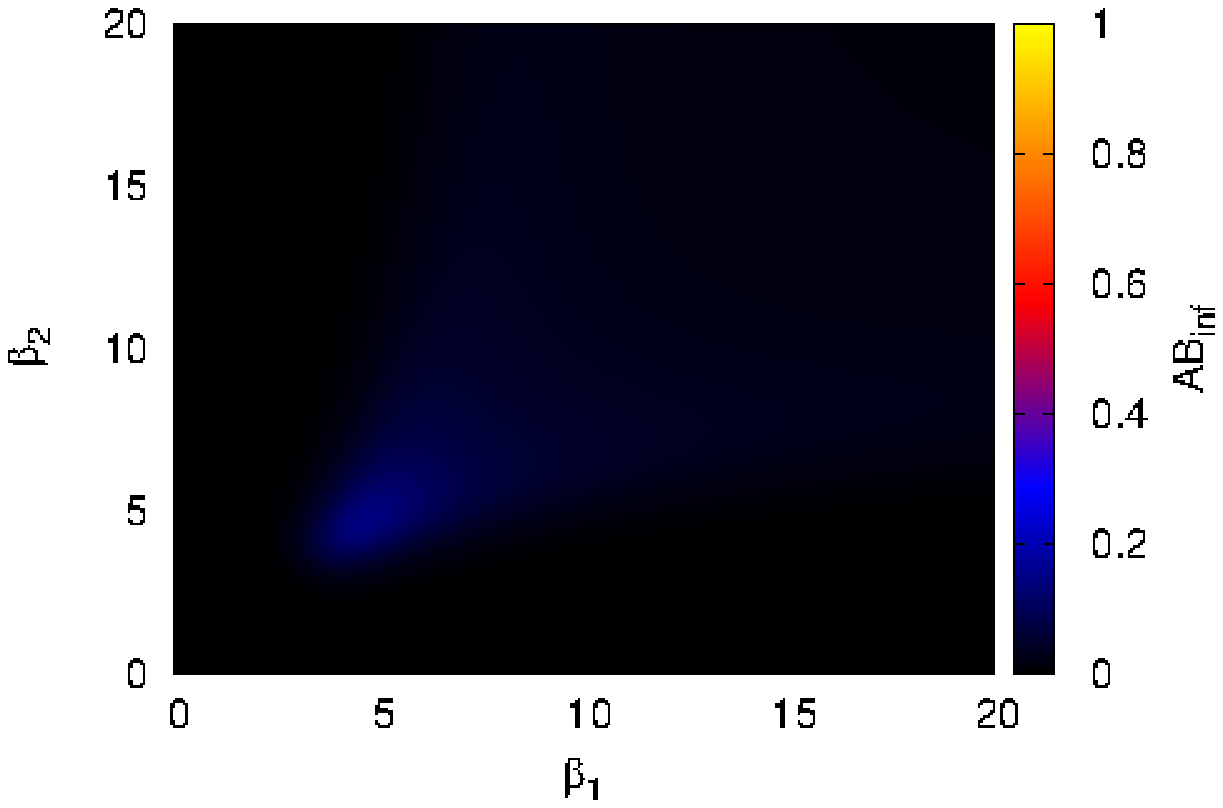}
   \label{fig:AB1-inf}
 }
 \subfigure[A=0.01, B=0.01]{
   \includegraphics[width=0.3\linewidth, height=2 in]{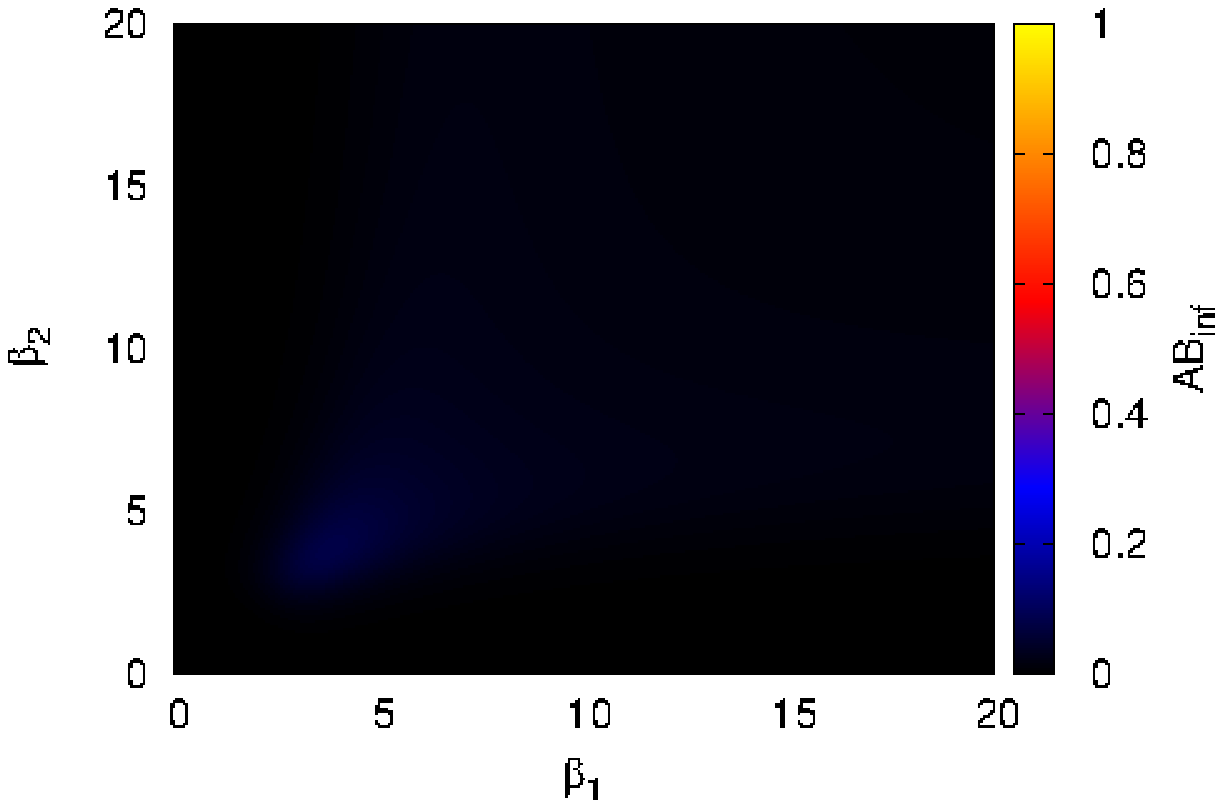}
   \label{fig:AB2-inf}
 }
\label{AB_infinity}
\caption{This shows how fraction of informed nodes with information 1 and 2 simultaneously changes with values of $\beta_1$ and $\beta_2$ as we change initial values of $A$ and $B$.}
\end{figure*}

\begin{figure*}[ht!]
\centering
\subfigure[A=0.0005, B=0.0005]{
   \includegraphics[width=0.3\linewidth, height=2 in] {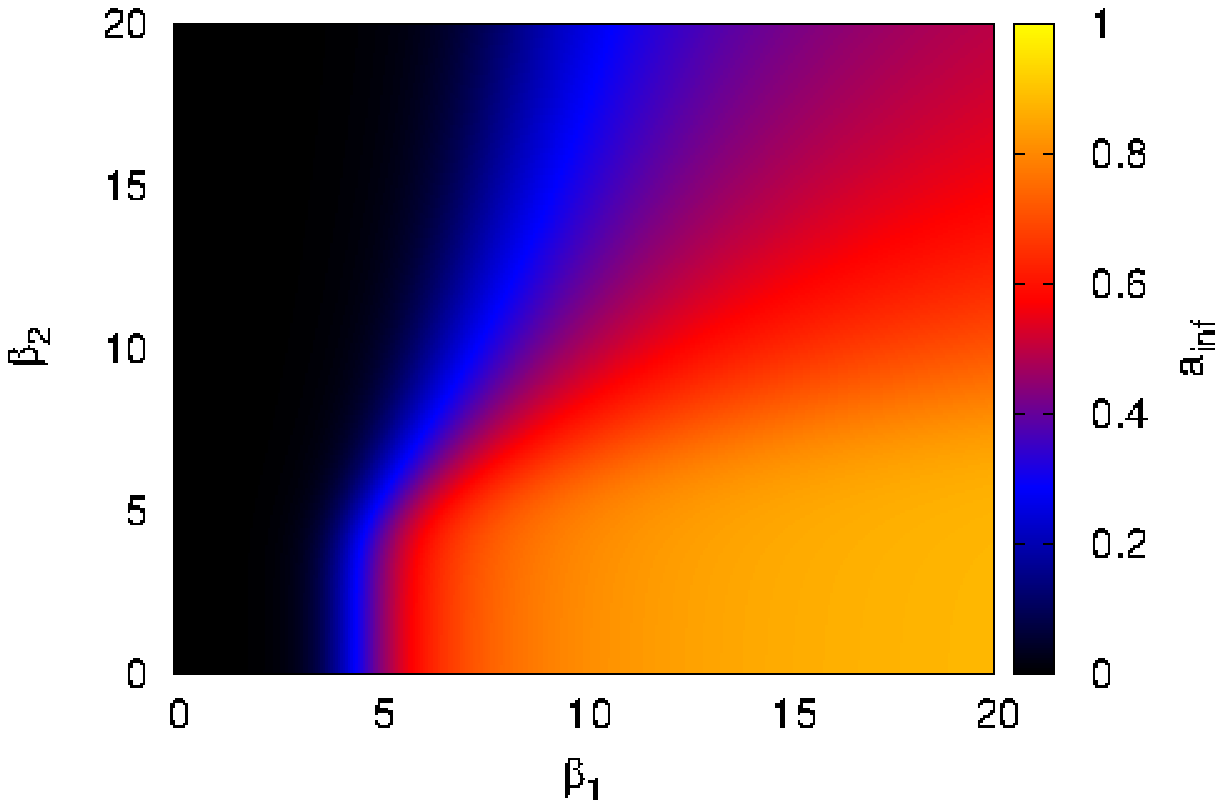}
   \label{fig:a-inf}
 }
\subfigure[A=0.001, B=0.001]{
   \includegraphics[width=0.3\linewidth, height=2 in] {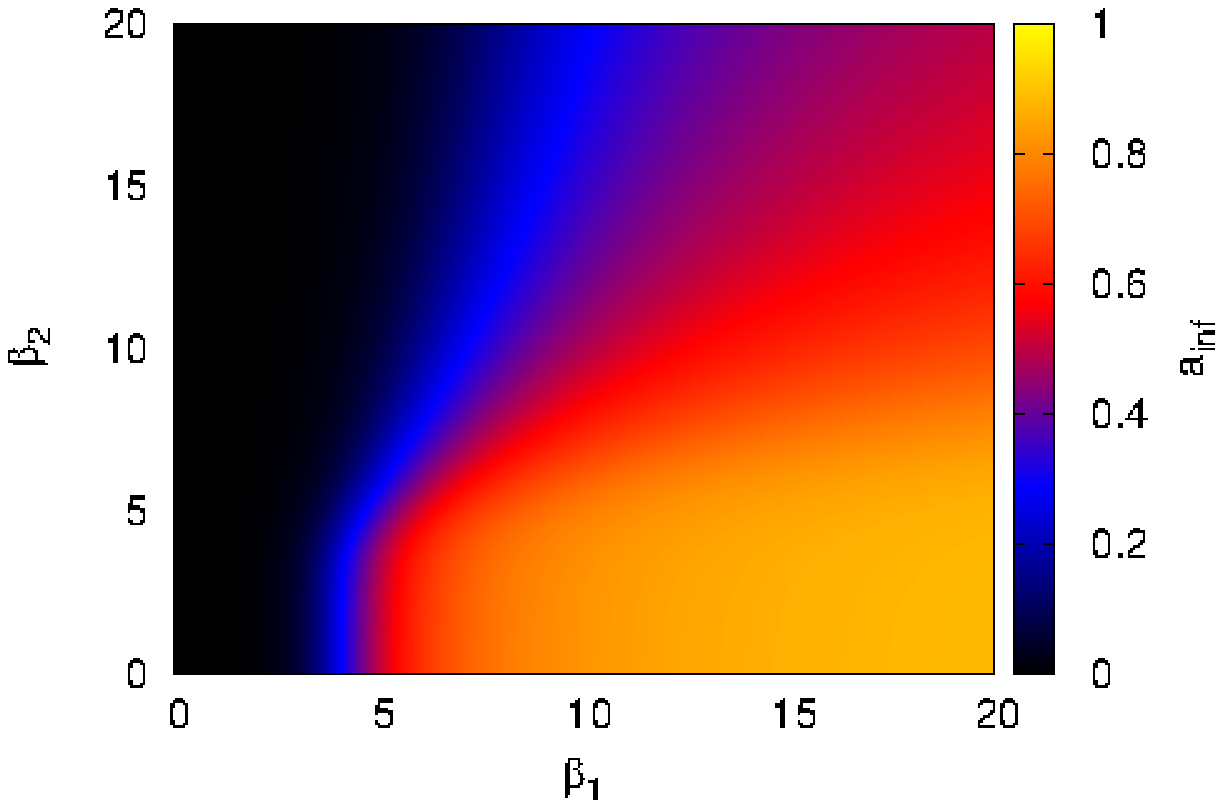}
   \label{fig:a1-inf}
 }
 \subfigure[A=0.01, B=0.01]{
   \includegraphics[width=0.3\linewidth, height=2 in]{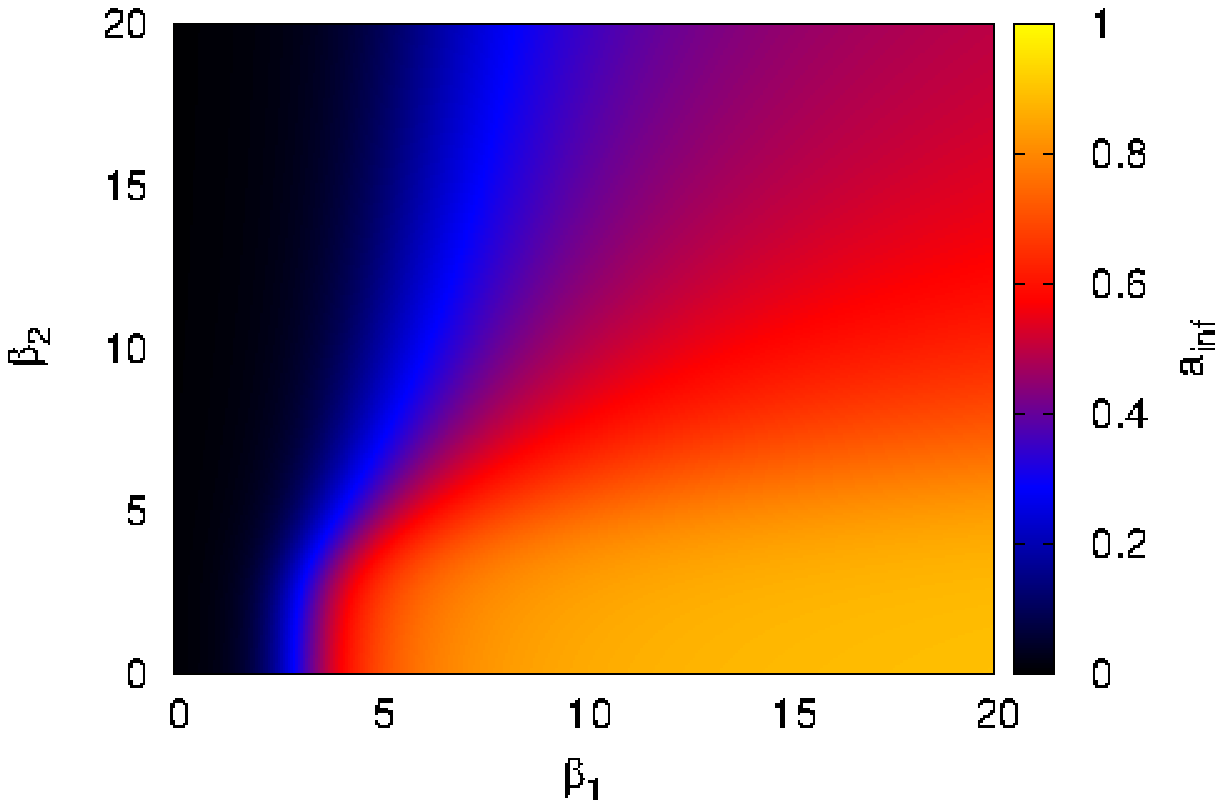}
   \label{fig:a2-inf}
 }
\label{a_infinity}
\caption{This shows how fraction of supporter nodes for information 1 changes for different initial values of A and B.}
\end{figure*}

\begin{figure*}[ht!]
\centering
\subfigure[A=0.0005, B=0.0005]{
   \includegraphics[width=0.3\linewidth, height=2 in] {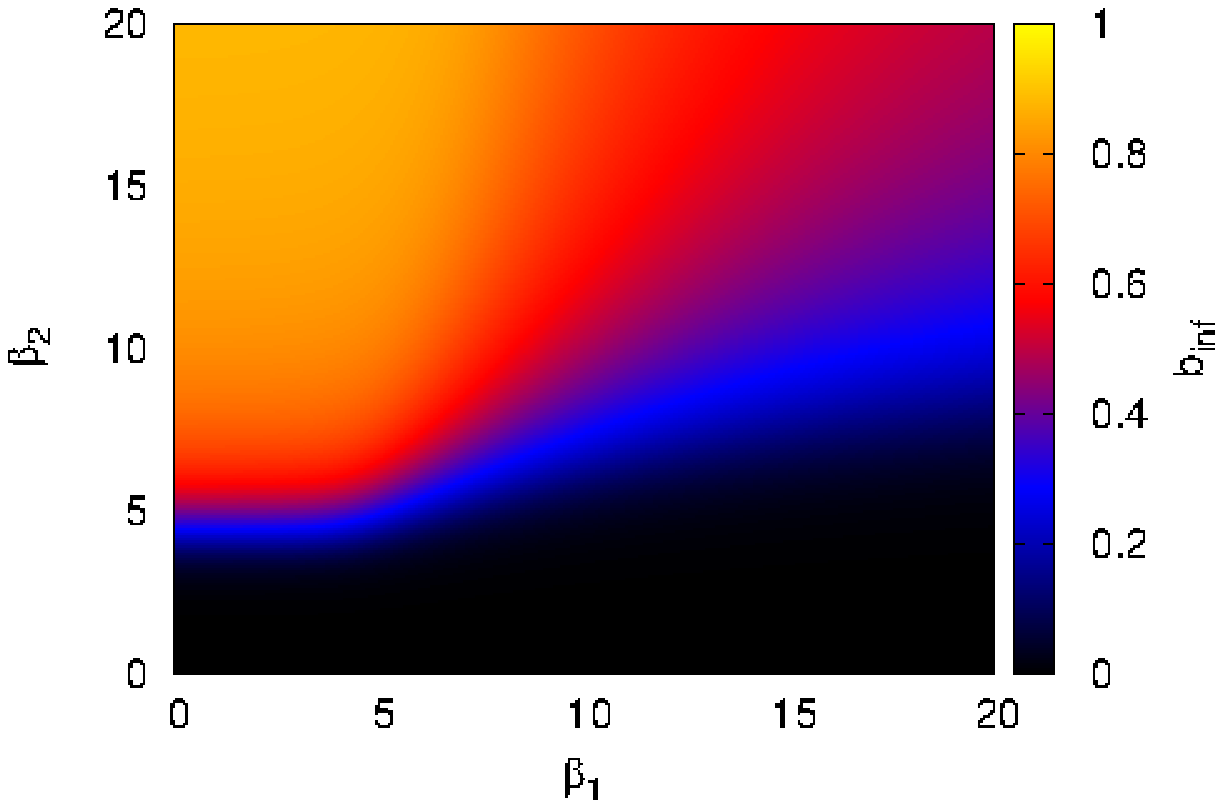}
   \label{fig:b-inf}
 }
\subfigure[A=0.001, B=0.001]{
   \includegraphics[width=0.3\linewidth, height=2 in] {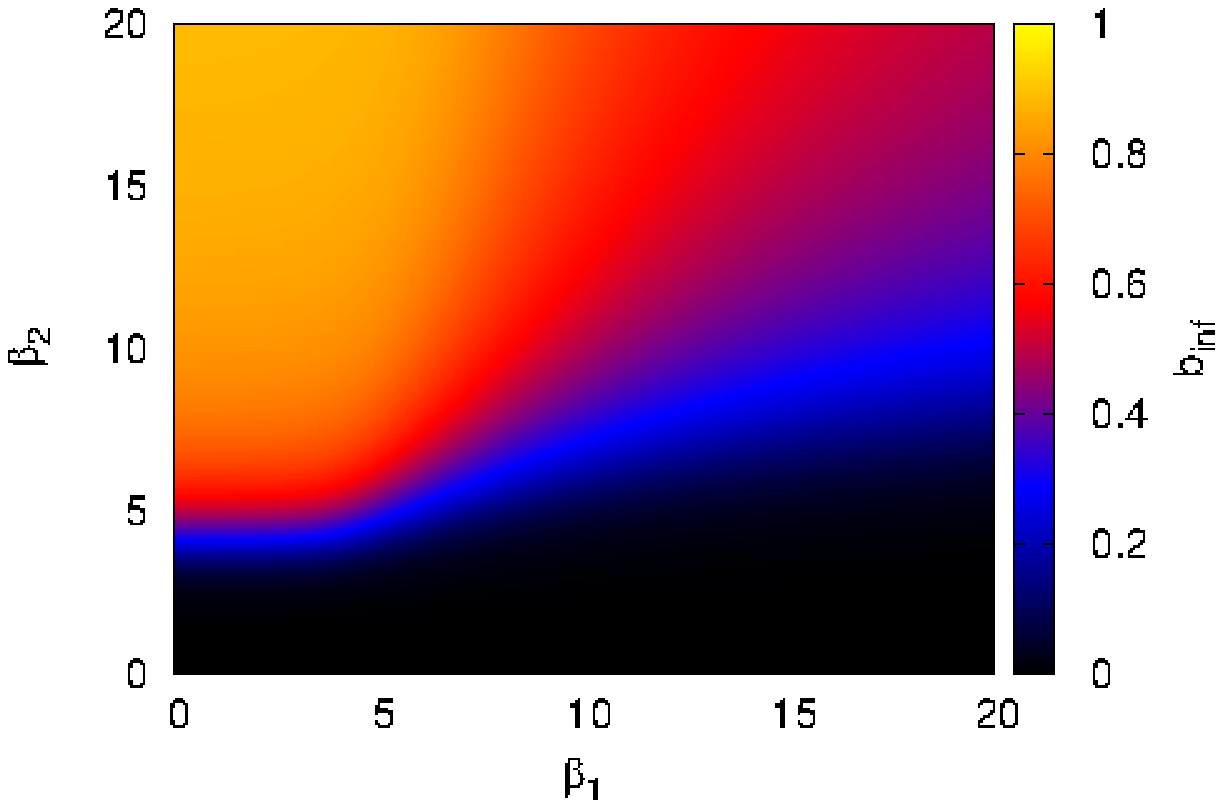}
   \label{fig:b1-inf}
 }
 \subfigure[A=0.01, B=0.01]{
   \includegraphics[width=0.3\linewidth, height=2 in]{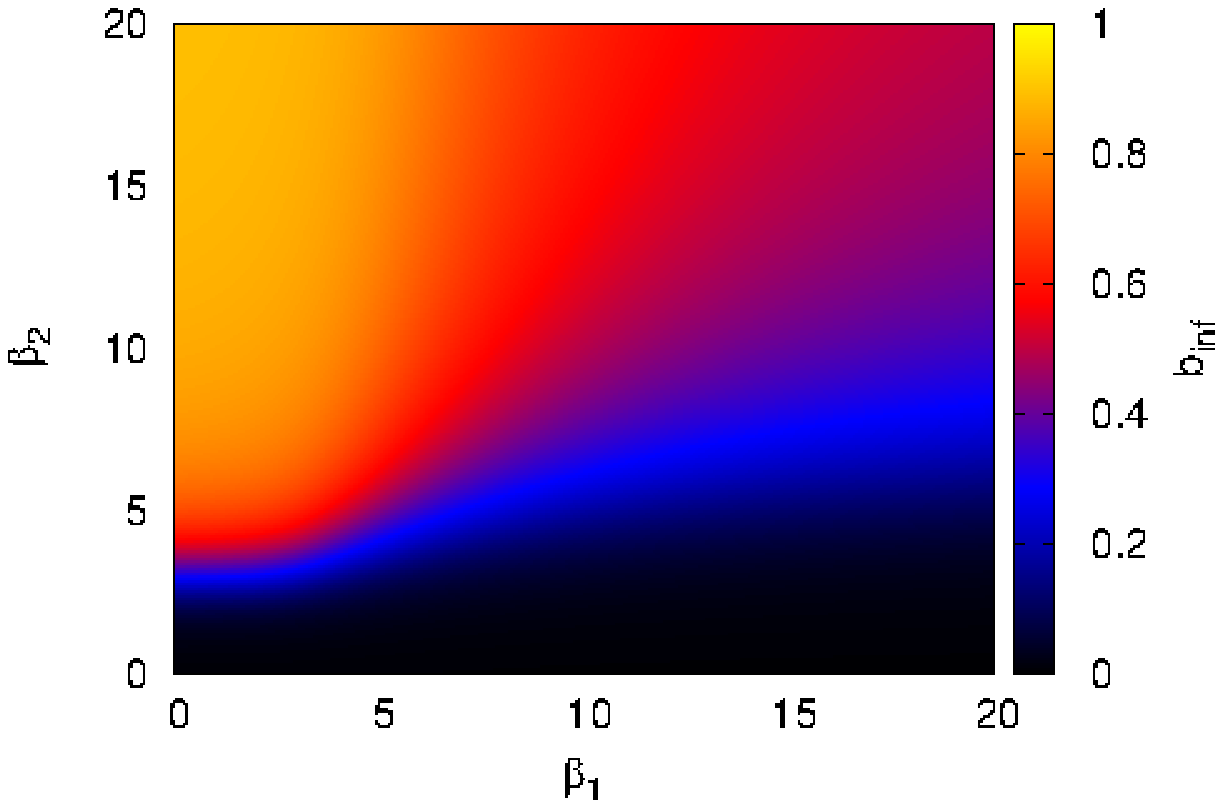}
   \label{fig:b2-inf}
 }
\label{fig:b-infinity}
\caption{This shows how fraction of supporter nodes for information 2 changes for different initial values of A and B.}
\end{figure*}
\clearpage
\section{Influence Calculation}
\label{app:InfluenceCalculation}
\begin{figure}[htb!]
\begin{center}
$\begin{array}{cc}
\includegraphics[width=0.45\linewidth, height=1.5 in]{Figures/icm4_12.eps} &
\includegraphics[width=0.45\linewidth, height=1.5 in]{Figures/icm3_12.eps}\\
\mbox{\textbf{(a) Information 1} } & \mbox {\textbf{(b) Information 2}}
\end{array}$
\end{center}
\caption{Influence Calculation in the network.} \label{fig:infoCal}
\vspace{-1em}
\end{figure}

Lets solve the value of $\alpha$ level by level for information 1. Hence, i=1.

At level 0, there is node 2 only which is also the seed node for information 1. Hence,
$\alpha^2_1$=1.

At level 1, there are node 1, 3 and 4. All these nodes are children of node 2. In these nodes, node 3 and 4 are also connected to each other.

For node 1, $$\alpha^1_1=\frac{\alpha^2_1}{d_1} + 0 =\frac{1}{1}=1.\ (0\ because\ no\ sibling\ node\ connection.)$$
For node 3, $$\alpha^3_1=\frac{\alpha^2_1}{d_3} + \frac{\alpha^4_1}{d_3}=\frac{1}{3}+\frac{(1/3)}{3}=\frac{4}{9}$$
For node 4, $$\alpha^4_1=\frac{\alpha^2_1}{d_4} + \frac{\alpha^3_1}{d_4}=\frac{1}{3}+\frac{(1/3)}{3}=\frac{4}{9}$$

At level 2, there are node 7 and 10. Node 7 is child of node 3 and node 10 is child of node 4.
For node 7, $$\alpha^7_1=\frac{\alpha^3_1}{d_7} + \frac{\alpha^{10}_1}{d_7}=\frac{(4/9)}{5}+\frac{((4/9)/2)}{5}=\frac{4}{30}$$
For node 10, $$\alpha^{10}_1=\frac{\alpha^4_1}{d_{10}} + \frac{\alpha^7_1}{d_{10}}=\frac{(4/9)}{2}+\frac{((4/9)/5)}{2}=\frac{4}{15}$$

At level 3, there are node 8 and 9. Node 8 and 9 are child of node 7.
For node 8, $$\alpha^8_1=\frac{\alpha^7_1}{d_8} + 0=\frac{(4/30)}{1}=\frac{4}{30}$$
For node 9, $$\alpha^9_1=\frac{\alpha^7_1}{d_9} + 0=\frac{(4/30)}{1}=\frac{4}{30}$$

Similarly, solve the value of $\alpha$ level by level for information 2. Hence, i=2.

At level 0, there is node 5 only which is also the seed node for information 2. Hence,
$\alpha^5_2$=1.

At level 1, there are node 6 and 7. All these nodes are children of node 5.

For node 6, $$\alpha^6_2=\frac{\alpha^5_2}{d_6} + 0 =\frac{1}{1}=1\ (0\ because\ no\ sibling\ node\ connection.)$$
For node 7, $$\alpha^7_2=\frac{\alpha^5_2}{d_7} + 0 =\frac{1}{5}$$

At level 2, there are node 3, 8, 9 and 10. All these nodes are children of node 7.

For node 3, $$\alpha^3_2=\frac{\alpha^7_2}{d_3} + 0 =\frac{(1/5)}{3}=\frac{1}{15}$$
For node 8, $$\alpha^8_2=\frac{\alpha^7_2}{d_8} + 0 =\frac{(1/5)}{2}=\frac{1}{5}$$
For node 9, $$\alpha^9_2=\frac{\alpha^7_2}{d_9} + 0 =\frac{(1/5)}{1}=\frac{1}{5}$$
For node 10, $$\alpha^{10}_2=\frac{\alpha^7_2}{d_{10}} + 0 = \frac{(1/5)}{2}=\frac{1}{10}$$

At level 3, there is only one node i.e., node 4. Node 4 has two parents.
For node 4, $$\alpha^4_2=\frac{(\alpha^3_2+\alpha^{10}_2)}{d_4} + 0 =\frac{(1/15)+(1/10)}{3}=\frac{1}{18}$$

\end{document}